\documentclass[prb,superscriptaddress,twocolumn]{revtex4-1}
\usepackage{times,amsmath}
\usepackage{epsfig}
\usepackage{color}
\usepackage{longtable}
\usepackage{graphicx}
\usepackage{dcolumn}
\usepackage{bm}
\usepackage{tabularx}
\usepackage{hyperref}
\usepackage{multirow}
\usepackage{appendix}
\hypersetup{citecolor=black, filecolor= black, linkcolor= black, urlcolor= black}

\begin{document}
\title{Structure-driven intercalated architecture of septuple-atomic-layer $MA_2Z_4$ family with diverse
\\ properties from semiconductor to topological insulator to Ising superconductor}

\author{Lei Wang}
\affiliation{Shenyang National Laboratory for Materials Science,
Institute of Metal Research, \\Chinese Academy of Science, 110016
Shenyang, Liaoning, P. R. China} 
\affiliation{School
of Materials Science and Engineering, University of Science and
Technology of China,\\Shenyang 110016, P. R. China} 

\author{Yongpeng Shi}
\affiliation{Shenyang National Laboratory for Materials Science,
Institute of Metal Research, \\Chinese Academy of Science, 110016
Shenyang, Liaoning, P. R. China} 
\affiliation{School
of Materials Science and Engineering, University of Science and
Technology of China,\\Shenyang 110016, P. R. China} 

\author{Mingfeng Liu}
\affiliation{Shenyang National Laboratory for Materials Science,
Institute of Metal Research, \\Chinese Academy of Science, 110016
Shenyang, Liaoning, P. R. China} 
\affiliation{School
of Materials Science and Engineering, University of Science and
Technology of China,\\Shenyang 110016, P. R. China} 

\author{Yi-Lun Hong}
\affiliation{Shenyang National Laboratory for Materials Science,
Institute of Metal Research, \\Chinese Academy of Science, 110016
Shenyang, Liaoning, P. R. China} 
\affiliation{School
of Materials Science and Engineering, University of Science and
Technology of China,\\Shenyang 110016, P. R. China} 

\author{Ming-Xing Chen}
\affiliation{School of Physics and Electronics, Hunan Normal
University,
Changsha 410081,
P. R. China} 

\author{Ronghan Li}
\affiliation{Shenyang National Laboratory for Materials Science,
Institute of Metal Research, \\Chinese Academy of Science, 110016
Shenyang, Liaoning, P. R. China} 
\affiliation{School
of Materials Science and Engineering, University of Science and
Technology of China,\\Shenyang 110016, P. R. China} 

\author{Qiang Gao}
\affiliation{Shenyang National Laboratory for Materials Science,
Institute of Metal Research, \\Chinese Academy of Science, 110016
Shenyang, Liaoning, P. R. China} 
\affiliation{School
of Materials Science and Engineering, University of Science and
Technology of China,\\Shenyang 110016, P. R. China} 

\author{Wencai Ren}
\affiliation{Shenyang National Laboratory for Materials Science,
Institute of Metal Research,
\\Chinese Academy of Science, 110016 Shenyang, Liaoning, P. R. China} 
\affiliation{School of Materials Science and
Engineering, University of Science and Technology of
China,\\Shenyang 110016, P. R. China} 

\author{Hui-Ming Cheng}
\affiliation{Shenyang National Laboratory for Materials Science,
Institute of Metal Research,
\\Chinese Academy of Science, 110016 Shenyang, Liaoning, P. R. China} 
\affiliation{School of Materials Science and
Engineering, University of Science and Technology of
China,\\Shenyang 110016, P. R. China} 
\affiliation{Shenzhen Geim Graphene Center, Tsinghua-Berkeley Shenzhen Institute (TBSI), \\ Tsinghua University, 1001 Xueyuan Road, Shenzhen 518055, P. R. China} 

\author{Yiyi Li}
\affiliation{Shenyang National Laboratory for Materials Science,
Institute of Metal Research,
\\Chinese Academy of Science, 110016 Shenyang, Liaoning, P. R. China} 
\affiliation{School of Materials Science and
Engineering, University of Science and Technology of
China,\\Shenyang 110016, P. R. China} 

\author{Xing-Qiu Chen}
\email{xingqiu.chen@imr.ac.cn} \affiliation{Shenyang National
Laboratory for Materials Science, Institute of Metal Research,
\\Chinese Academy of Science, 110016 Shenyang, Liaoning, P. R. China} 
\affiliation{School of Materials Science and
Engineering, University of Science and Technology of
China,\\Shenyang 110016, P. R. China} 

\date{\today}
\begin{abstract}
Motivated by the fact that septuple-atomic-layer MnBi$_2$Te$_4$ can
be structurally viewed as the combination of double-atomic-layer
MnTe intercalating into quintuple-atomic-layer Bi$_2$Te$_3$, we
present a general approach of constructing twelve
septuple-atomic-layer $\alpha_i$- and $\beta_i$-$MA_2Z_4$ monolayer
family (\emph{i} = 1 to 6) by intercalating MoS$_2$-type $MZ$$_2$
monolayer into InSe-type A$_2$Z$_2$ monolayer. Besides reproducing
the experimentally synthesized $\alpha_1$-MoSi$_2$N$_4$,
$\alpha_1$-WSi$_2$N$_4$ and $\beta_5$-MnBi$_2$Te$_4$ monolayer
materials, another 66 thermodynamically and dynamically stable
$MA_2Z_4$ were predicted, which span a wide range of properties upon
the number of valence electrons (VEC). $MA_2Z_4$ with the rules of
32 or 34 VEC are mostly semiconductors with direct or indirect band
gap and, however, with 33 VEC are generally metal, half-metal
ferromagnetism, or spin-gapless semiconductor upon whether or not an
unpaired electron is spin polarized. Moreover, we propose
$\alpha_2$-WSi$_2$P$_4$ for the spin-valley polarization,
$\alpha_1$-TaSi$_2$N$_4$ for Ising superconductor and
$\beta_2$-SrGa$_2$Se$_4$ for topological insulator.
\end{abstract}

\maketitle

\begin{figure*}
\begin{center}
\includegraphics[width=0.82\textwidth]{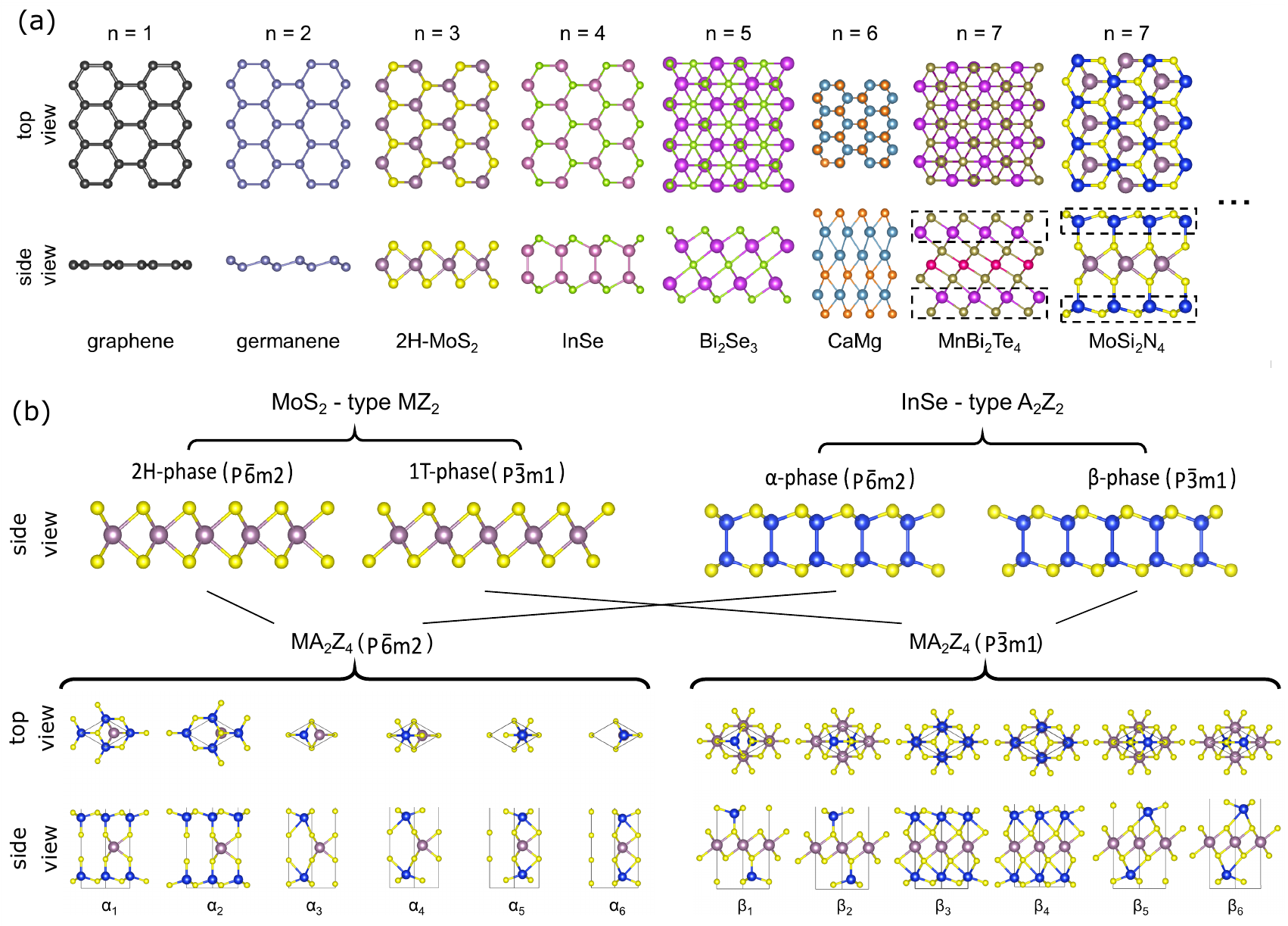}
\caption{(Color online) Panel (a), Representatives for 2D materials
with increasing atomic-layer number, ``\emph{n}''. Panel (b), twelve
possible constructions ($\alpha_i$ and $\beta_i$ \emph{n} = 7
monolayer structure, \emph{i} = 1, to 6) by intercalating
MoS$_2$-type $MZ_2$ \emph{n} = 3 monolayer into broken InSe-type
$A_2Z_2$ \emph{n}=4 monolayer. Note that all $\alpha_i$ and
$\beta_i$ \emph{n} = 7 monolayer structures share the same space
groups of \emph{P}$\overline{6}m$2 and \emph{P}$\overline{3}m$1,
respectively.} \label{model}
\end{center}
\end{figure*}

\begin{figure*}
\begin{center}
\includegraphics[width=0.82\textwidth]{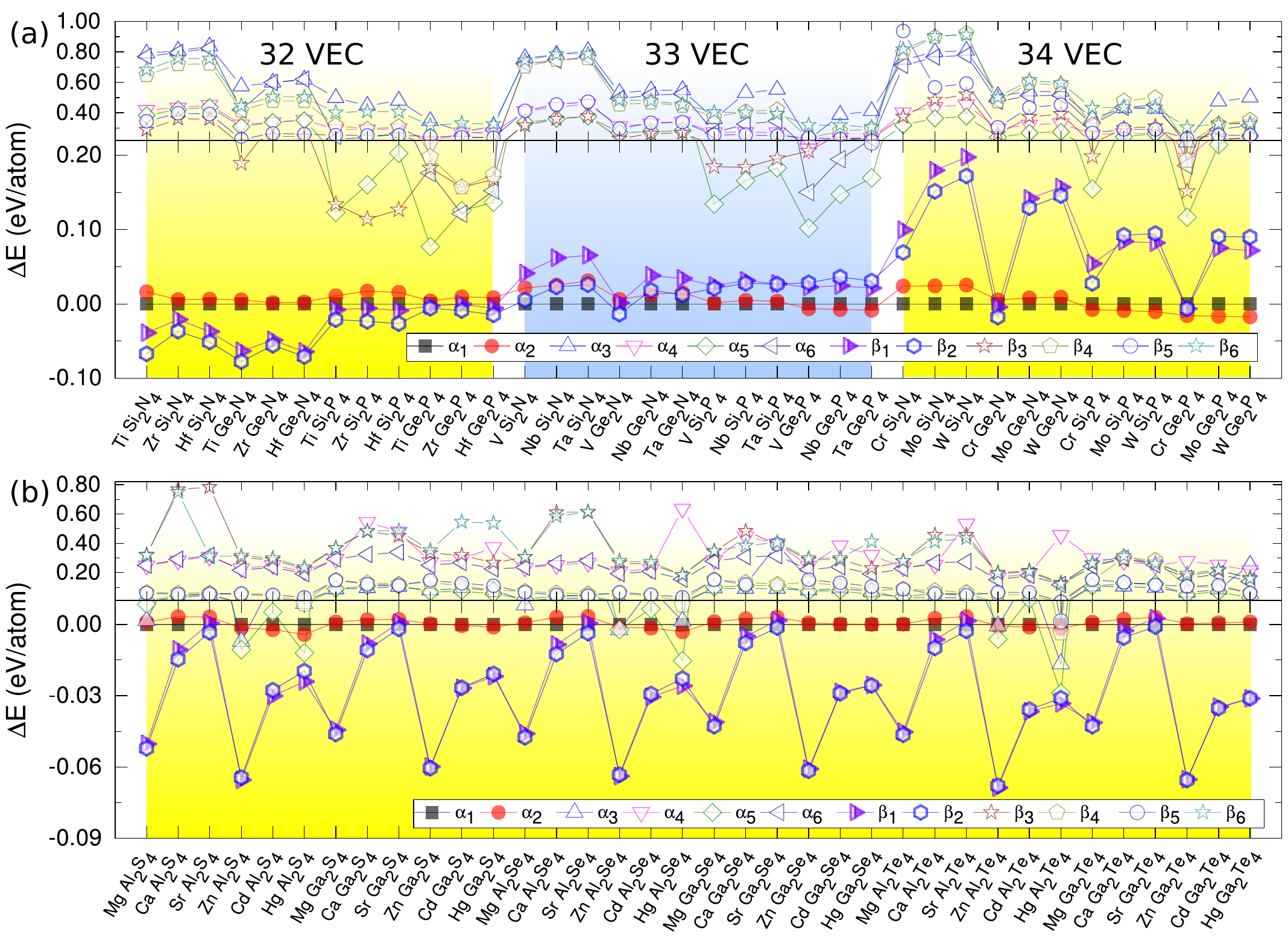}
\caption{The DFT-derived enthalpies of formation of 12 competing
structural candidates with respect to that of $\alpha_1$ candidate
for 36 $MA_2Z_{4}$ monolayer materials with \emph{M} = first
transition metal elements with 32, 33, and 34 VECs in panel (a) and
alkali earth elements with 32 VEC in panel (b), respectively.}
\label{energy}
\end{center}
\end{figure*}

\begin{figure}
\begin{center}
\includegraphics[width=0.45\textwidth]{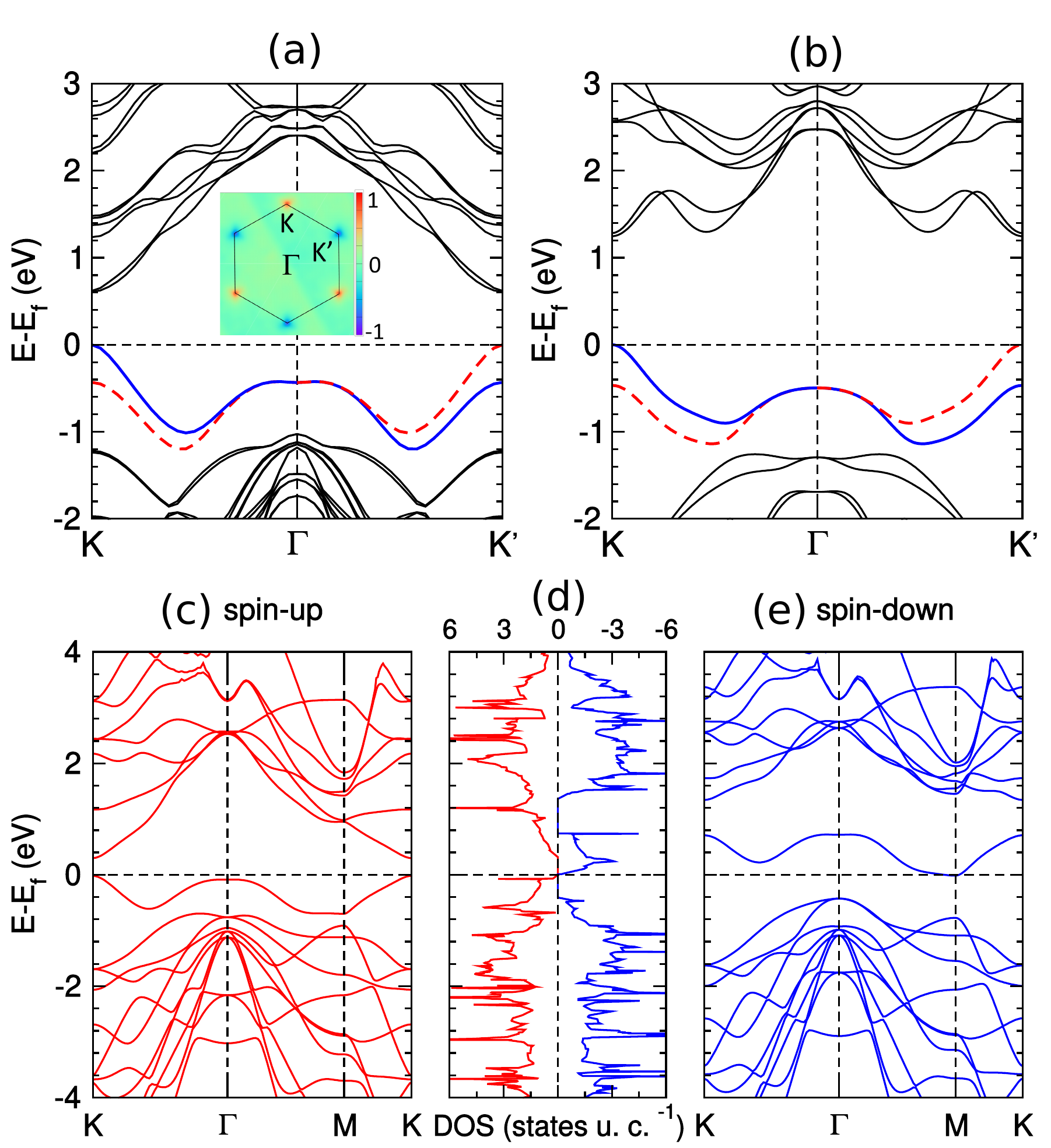}
\caption{(color online) (a) Spin-valley coupling of
$\alpha_2$-WSi$_2$P$_4$ in comparison with 2\emph{H}-WSe$_2$ in
(b).Inset of panel (a): Derived Berry curvatures of
$\alpha_2$-WSi$_2$P$_4$. (c and e) the calculated electronic band
structures for the majority spin-up and minority spin-down channels
of the spin-gapless $\alpha_1$-VSi$_2$P$_4$ monolayer semiconductor,
along together with their corresponding electronic densities of
states in panel (d).
 } \label{fig3}
\end{center}
\end{figure}

Due to the potential applications in energy storage and conversion
~\cite{graphene_Science_2015,graphene_NM_2015}, nanoelectronics
~\cite{akinwande_2D_2014,chang_large-area_2016}, and spintronics
~\cite{PRL-MoS2-spintronic,pesin_spintronics_2012}, as well as
superconductivity~\cite{
Type-II-PRL-wang,Lu1353,Ising-Zhou-PRB,saito_NPs_2016,Inte-Liu-PRX,
xi_ising_2016,de_tuning_2018,falson_type-ii_2020,profeta_phonon_2012,
si_first_2013}, two-dimensional (2D)
hexagonal monolayer materials have been attracting tremendous
interest in both experimental and theoretical studies, which is
inseparable from their rich geometric structures and chemical
compositions.

To dates, some 2D monolayer materials have been discovered while
their atomic-layer numbers are limited to just a few. In 2004, the
single atomic layer (\emph{n} = 1) graphite in Fig.~\ref{model}(a),
namely graphene, was experimentally realized by mechanical
exfoliation method, giving rise to the birth of 2D materials.
Graphene is a semimetal with the occurrence of massless Dirac cone
due to the $\sigma$ bonding hopping, leading to a few special
properties, {\it e.g.} ultra-high carrier
mobility~\cite{banszerus_mobility_2015}, high mechanical
strength~\cite{lee_measurement_2008}, high thermal
conductivity~\cite{pop_thermal_2012} and high optical
transparency~\cite{sheehy_optical_2009,kuzmenko_universal_2008}. In
similarity to graphene from graphite, one-atomic-layer $h$-BN
monolayer can also be exfoliated from its bulk form, but with a wide
gap of about 6 eV
\cite{h-BN_NM_2004,song_large_2010}. When two
carbon atoms of graphene in its unit cell are replaced by Si, Ge or
Sn atom, its flat $n$ = 1 monolayer structure will be slightly
buckled into a double-atomic-layer \emph{n} = 2 monolayer structure
in Fig.~\ref{model}(a). When $n$ comes to three, monolayer
transition metal dichalcogenides
(TMDCs)\cite{PRL-MoS2,wang_electronics_2012} become highly rich in
both compositions and properties, spanning a wide range from
semimetals, semiconductors, and to superconductors as well as to
topological insulators. For example, \emph{n} = 3 2$H$-WTe$_{2}$
monolayer (Fig.~\ref{model}(a)) is predicted to be a weyl semimetal
~\cite{soluyanov_type-ii_2015,li_evidence_2017} with an anomalous
giant magnetoresistance and
superconductivity~\cite{eftekhari_tungsten_2017}, while that its
distorted 1\emph{T} monolayer structure is predicted to be a quantum
spin Hall (QSH) insulator~\cite{xiang_quantum_2016}. Another type of
\emph{n} = 3 monolayer material is the recently discovered van der
Waals (vdW) 2D ferromagnetic semiconductor CrI$_3$, with a very
large tunneling magnetoresistance, of which the magnetism can be
manipulated by the bias electric field and electrostatic
doping~\cite{jiang2018controlling,huang2018electrical,wang2018very}.
Monolayer InSe (Fig.~\ref{model}(a)) of group-III monochalcogenides
\cite{li_2015_NR} consisting of quadruple-atomic-layer \emph{n} = 4
monolayer can be used in photocatalyst \cite{zhuang_CM_2013} and the
hole-doped monolayer InSe even has a strong electron-phonon
coupling, which affects its transport and optical
properties\cite{sun_inse_2018,lugovskoi_strong_2019}. As to the
quintuple-atomic-layer \emph{n} = 5 monolayer structure
(Fig.~\ref{model}(a)), Bi$_{2}$Se$_{3}$ is a famous case of
topological insulator (TI)~\cite{zhang_topological_2009,hsieh_tunable_2009}.
Another \emph{n} = 5 monolayer
CrGeTe$_3$ is also a vdW 2D magnet with the potential to be applied
in ultra-compact spintronics ~\cite{gong2017discovery}. Furthermore,
the known sextuple-atomic-layer \emph{n} = 6 monolayer material
(\emph{e.g.}, CaMg in Fig.~\ref{model}(a)) was theoretically
predicted in the reported 2DMatPedia
database~\cite{zhou_2dmatpedia_2019}. With further increasing atomic
layer $n$, the vdW MnBi$_{2}$Te$_{4}$ and septuple-atomic-layer
\emph{n}=7  MnBi$_2$Te$_4$ monolayer (Fig.~\ref{model}(a)) were
reported to be the antiferromagnetic and ferromagnetic topological
insulators, respectively\cite{zhang_topological_2019,
li_intrinsic_2019,otrokov_prediction_2019,rienks_large_2019}, which
naturally possesses anomalously quantum spin Hall effect
\cite{deng_quantum_2020}. Most recently, another type of $n$ = 7
monolayer material of MoSi$_{2}$N$_{4}$ (Fig.~\ref{model}(a)) has
been reported, which is a semiconductor with a band gap of about
1.94 eV~\cite{MSN}. As illustrated in Fig.~\ref{model}(a), these
known monolayer 2D materials consisting of \emph{n} = 1, 2, 3, 4, 5,
6, and 7 atomic layer thicknesses have attracted tremendous interest
for their structures, physics and potential applications. Certainly,
there is no doubt that, with varying \emph{n} number, compositions
and constituents, they will become richer in both structures and
properties. However, the difficulties lie in how we effectively seek
for more monolayer materials with promising properties.

Within this context, we have proposed a general intercalated
architecture approach to systemically construct \emph{n}=7 $MA_2Z_4$
monolayer family on basis of first-principles density functional
theory. Besides reproducing the experimentally synthesized
$\alpha_1$-MoSi$_2$N$_4$, $\alpha_1$-WSi$_2$N$_4$ and
$\beta_5$-MnBi$_2$Te$_4$ monolayer materials, we predict 66,
thermodynamically and dynamically, stable $MA_2Z_4$ monolayer
materials with diverse properties, which can be classified via the
number of valence electrons (VEC). $MA_2Z_4$ with the rules of 32 or
34 VEC are mostly semiconductors with direct or indirect band gap
and, however, with 33 VEC are generally metal, half-metal
ferromagnetism, or spin-gapless semiconductor upon whether or not an
unpaired electron is spin polarized. Additionally, we suggest
$\alpha_2$-WSi$_2$P$_4$ monolayer material for the spin-valley
polarization, $\alpha_1$-TaSi$_2$N$_4$ for Ising superconductor and
$\beta_2$-SrGa$_2$Se$_4$ for topological insulator.

\section{Results}
\textbf{Intercalated architecture approach.} If we look back these
known 2D monolayer materials in Fig.~\ref{model}(a), they seem to
share a general scheme (here called \emph{intercalated
architecture}) to construct various 2D structures within atomistic
scale. A \emph{n}=7 MnBi$_{2}$Te$_{4}$ monolayer was viewed as the
(111) plane of rocksalt structure MnTe inserted into the \emph{n}=5
Bi$_{2}$Se$_{3}$
monolayer~\cite{zhang_topological_2009,hsieh_tunable_2009}. In a
sense, it seems inherit the topology of Bi$_{2}$Se$_{3}$ and the
magnetism of MnTe. By analyzing \emph{n}=7 MoSi$_{2}$N$_{4}$
monolayer, it can be viewed as the insertion of the \emph{n}=3
2\emph{H}-MoS$_2$-type MoN$_{2}$ monolayer into the \emph{n}=4
$\alpha$-InSe-type Si$_{2}$N$_{2}$
monolayer~\cite{Ozdamar_PRB_2018}. With such a special insertion and
structural stacking, monolayer MoSi$_2$N$_4$ seems to inherit the
semiconducting gap of $\alpha$-Si$_2$N$_2$ (1.74 eV derived by PBE
~\cite{Ozdamar_PRB_2018}) and interesting tunable properties from
MoS$_2$-type MoN$_2$. Importantly, MoS$_2$-type monolayer has two
frequently observed \emph{n}=3 2\emph{H} ($P$$\overline{6}$$m$2) and
1\emph{T} ($P$$\overline{3}$m1) monolayer structures, and InSe-type
monolayer usually crystallizes in two prototypical $n$=4 $\alpha$
($P$$\overline{6}$$m$2) and $\beta$ ($P$$\overline{3}$m1) monolayer
structures~\cite{Ozdamar_PRB_2018}. Hence, we elucidate this process
of intercalated architecture in Fig. ~\ref{model}(b) in a more
general way. Within the assumption of the same space group for two
intercalating monolayer units: (1) 2\emph{H}-MoS$_2$-type $MZ_2$
monolayer can be inserted into $\alpha$-InSe-type $A_2Z_2$ monolayer
within the same $P$$\overline{6}$$m$2 space group to form six
possible \emph{n}=7 $\alpha_i$-$MA_2Z_4$ monolayer structures
(\emph{i} = 1 to 6 in the left six panels of Fig.~\ref{model}(b));
(2) 1\emph{T}-MoS$_2$-type $MZ_2$ monolayer can be inserted into
$\beta$-$A_2Z_2$ monolayer within the same ($P$$\overline{3}$m1)
space group to also form the other six possible \emph{n}=7
$\beta_i$-$MA_2Z_4$ monolayer structures (\emph{i} = 1 to 6 in the
right six panels of Fig.~\ref{model}(b)). It needs to be emphasized
that these six $\alpha_i$ and six $\beta_i$ (\emph{i} = 1 - 6)
monolayer structures indeed connect to each other through mirror and
translation operations of double layer unit $AZ$, respectively.

As a benchmark of the structural modeling reliability, we have first
tested the first-principle structural optimizations (supplementary
materials~\cite{SM}) of three experimentally already-synthesized
\emph{n}=7 monolayer materials of
MnBi$_2$Te$_4$~\cite{zhang_topological_2019,
li_intrinsic_2019,otrokov_prediction_2019,rienks_large_2019},
MoSi$_2$N$_4$ and WSi$_2$N$_4$ ~\cite{MSN} by considering these 12
possible structural candidates. Our calculations reveal that the
$\alpha_1$-monolayer structure is energetically favorable for both
MoSi$_2$N$_4$ and WSi$_2$N$_4$, whereas the $\beta_5$ monolayer
structure is the most stable one for MnBi$_2$Te$_4$ monolayer. The
obtained structures for them are in perfect agreement with the known
experiments~\cite{zhang_topological_2019,
li_intrinsic_2019,otrokov_prediction_2019,rienks_large_2019}.

\textbf{Prediction of $MA_2Z_4$ family.} Furthermore, we have
extended our DFT structural optimizations by considering a large
number of \emph{n}=7 \emph{M}\emph{A}$_2$\emph{Z}$_4$ monolayer
family by varying atomic constituents ($M$ -- the transition metal
elements IVB, VB and VIB groups; $A$ -- IVA-group elements and $Z$
-- VA-group elements). Utilizing DFT calculations, we have performed
the structural optimizations by considering all 12 monolayer
candidates constructed above for each composition. As shown in
Fig.~\ref{energy}(a), we have compiled the relative enthalpies of
formations of \emph{M}\emph{A}$_2$Z$_4$ ($M$ = Ti, Zr, Hf, V, Nb,
Ta, Cr, Mo, W; $A$ = Si and Ge; $Z$ = N and P) with respect to their
$\alpha_1$ candidate. Interestingly, we have found that their
stabilities in energy seem to be correlated with the number of the
valence electrons per formula unit (VEC). $MA_2Z_4$ monolayer with
32 VEC are all stable at its $\beta_2$ phase in
Fig.~\ref{energy}(a), but in the series of 33 or 34 VEC the
structural stabilities become a bit complicated. \emph{M}Si$_2$N$_4$
(\emph{M} = V, Nb, Ta, Cr, Mo, W), \emph{M}Ge$_2$N$_4$ (\emph{M} =
Nb, Ta, Mo, W), and $M$Si$_2$P$_4$ (\emph{M} =V, Nb, Ta) are
energetically the lowest in their $\alpha_1$ monolayer structure,
whereas \emph{M}Si$_2$P$_4$ (\emph{M} =Cr, Mo, W) and
\emph{M}Ge$_2$P$_4$ ($M$ = V, Nb, Ta, Cr, Mo, W) are energetically
most favorable in their $\alpha_2$ monolayer phase. The obtained
absolute enthalpies of formation and their corresponding optimized
lattice constants and structural parameters are further compiled in
supplementary Table S1 and S2~\cite{SM}, respectively. Moreover, we
have calculated their phonon dispersions of all these 36 compounds
in supplementary Fig. S1~\cite{SM}. Among them, 32 compounds are
stable, both dynamically and thermodynamically, and only 4 compounds
($\beta_2$-TiSi$_2$N$_4$, $\beta_2$-TiGe$_2$N$_4$,
$\beta_2$-VGe$_2$N$_4$ and $\beta_2$-CrGe$_2$N$_4$) are dynamically
unstable, due to their imaginary phonon dispersions.

\begin{figure*}
\begin{center}
\includegraphics[width=0.78\textwidth]{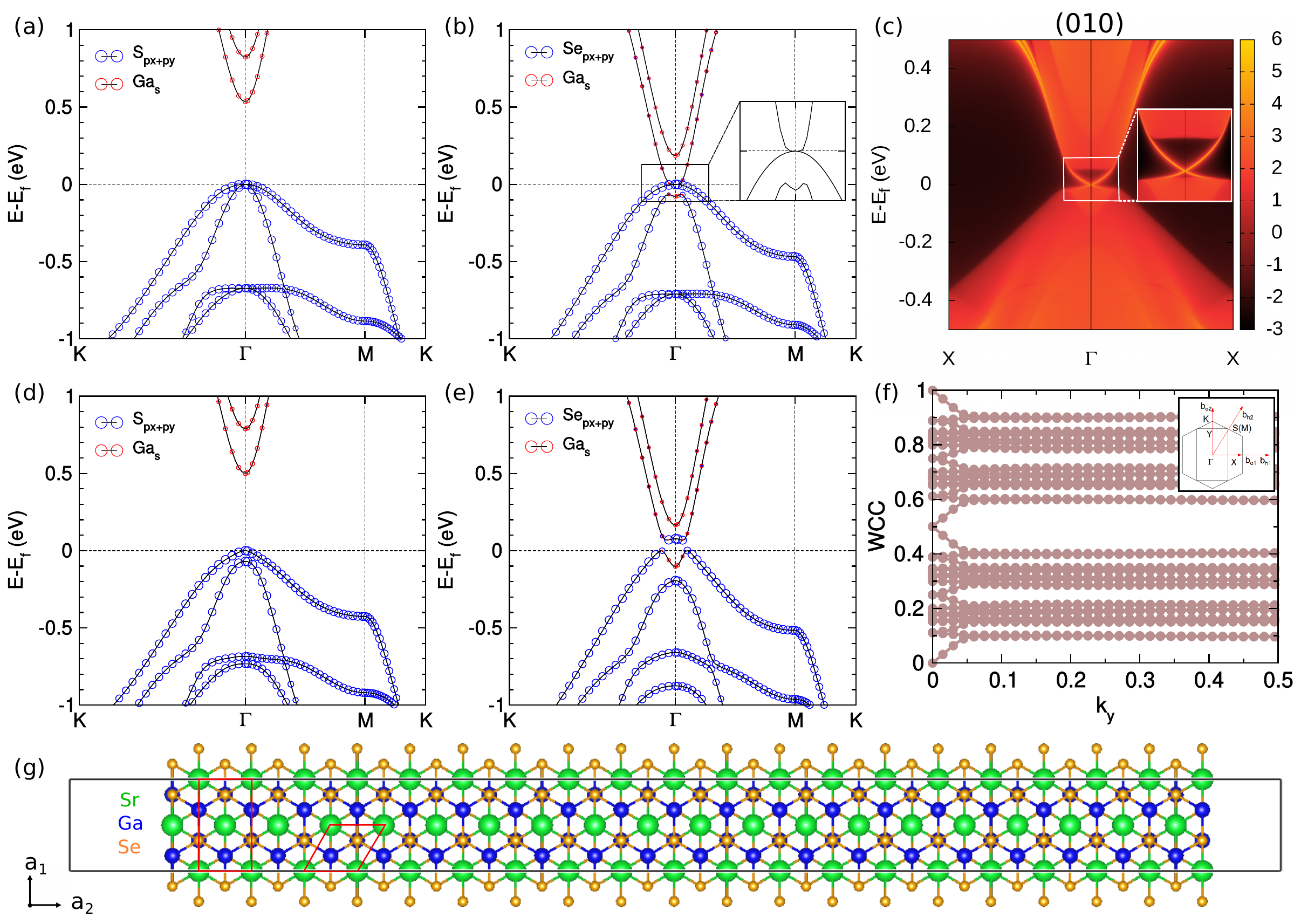}
\caption{(Color online) Panel (a),(b) and (d), (e) are electronic band
structures without and with including SOC of $\beta_2$-SrGa$_2$S$_4$ and
$\beta_2$-SrGa$_2$Se$_4$, respectively. The insert in Panel (b) shows band
degeneracy at Fermi level. Panel (c) is edge states of (010)
edge of orthogonal cell and the insert shows the linear dispersive edge
states. Panel (f) is evolution of  Wannier charge center (WCC) in the
k$_z$ = 0 plane, which implies a nonzero topological invariant. The
insert in Panel (f) is the BZ of hexagonal cell and orthogonal cell.
Panel (g) is the top view of
20-cell $\beta_2$-SrGa$_2$Se$_4$ nanoribbon with (010) edge, where
the red solid-line square and rhombus are orthogonal cell and hexagonal
cell, respectively.} \label{topo}
\end{center}
\end{figure*}

In addition, we have paid our attention to the other 36 $MA_2Z_4$
monolayer materials with 32 VEC (\emph{M} = alkali earth elements
(Mg, Ca, Sr) and the IIB-group elements (Zn, Cd, Hg), \emph{A} = Al
and Ga, and \emph{Z} = S, Se, and Te). As shown in
Fig.~\ref{energy}(b), all these monolayer materials crystallize in
the most stable $\beta_1$ or $\beta_2$ structures. Fortunately, we
have found that among these 36 predictions only
$\beta_1$-ZnAl$_2$S$_4$ was already mentioned in the reported
2DMatPedia database which were obtained through conventional
high-throughput computational method via both top-down and bottom-up
discovery procedures~\cite{zhou_2dmatpedia_2019}. This fact further
demonstrates the power and reliability of our currently proposed
intercalated architecture.

\textbf{Electronic structures.} We have derived their electronic
band structures for these selected 36 compounds in
Fig.~\ref{energy}(a) at their most stable structures in
supplementary Fig. S2~\cite{SM}. $MA_2$Z$_4$ monolayer with 32 VEC
are predicted to be semiconductor for all nitrogen-based compounds,
whereas is metallic for all phosphorus-based compounds. We have also
noted that, except for a magnetic CrGe$_2$N$_4$, $MA_2$N$_4$ with 34
VEC is also semiconductor. It can be understandable for the
occurrence of semiconductors, according to the ionic picture
satisfying the closed-shell electronic configuration of $M$$^{4+}$,
$A$$^{4+}$, and $Z$$^{3-}$ for 32 VEC for \emph{M} = IVB-group Ti,
Zr, and Hf elements or 34 VEC for $M$ = VIB-group Cr, Mo and W
elements due to a remained fully occupied $s^2$ orbital. Although
phosphorus-based compounds with 32 VEC also form closed-shell
electron configuration, they are metallic mainly because phosphorus
atom has a lower electronegativity than that of nitrogen. We have
summarized their band gaps of 17 semiconducting monolayer materials
in supplementary Table S2~\cite{SM}. Four materials of
$\beta_2$-ZrGe$_2$N$_4$ and $\beta_2$-HfGe$_2$N$_4$ and
$\alpha_2$-MoSi$_2$P$_4$ and $\alpha_2$-WSi$_2$P$_4$ are a direct
band-gap semiconductor. The other 13 compounds exhibit the indirect
band gaps mostly from the highest valence top at $\Gamma$ to the
lowest conduction bottom at \emph{M} and \emph{K} within the
framework of the conventional DFT calculations. Their direct band
gaps are estimated to be 1.04 eV at $\Gamma$, 1.15 eV at $\Gamma$,
0.91 eV at K and 0.86 eV at K for $\beta_2$-ZrGe$_2$N$_4$,
$\beta_2$-HfGe$_2$N$_4$, $\alpha_2$-MoSi$_2$P$_4$ and
$\alpha_2$-WSi$_2$P$_4$, respectively. Due to conventional DFT
problem to underestimate band gap, we have further used hybrid DFT
(HSE06) method to correct their band gaps to 2.34 eV, 2.45 eV, 1.19
eV and 1.11 eV, respectively. The case of $\alpha_1$-WSi$_2$N$_4$
has the largest indirect band gap of 2.08 eV (HSE: 2.66 eV).
Although within the framework of conventional DFT three compounds of
$\alpha_1$-CrSi$_2$N$_4$, $\alpha_2$-CrSi$_2$P$_4$ and
$\alpha_2$-WGe$_2$P$_4$ are indirect band-gap semiconductors, the
HSE calculations revise them to the appearance of direct band gaps
of 0.94 eV at K, 0.65 eV at K and 0.89 eV at K, respectively
(supplementary Table S2~\cite{SM}).

\textbf{VEC of 32 and 34.} Similar to the monolayer of
TMDCs~\cite{di_coupled_2012}, our materials of $MA_2$Z$_4$ with 32
or 34 VEC are also lacking of inversion symmetry with a strong spin
orbit coupling (SOC) effect from the heavy elements \emph{M}. Hence,
many of them are expected to exhibit rich spin-valley physics.
Taking $\alpha_2$-WSi$_{2}$P$_{4}$ as an example
(Fig.~\ref{fig3}(a)), the two valleys at $K$ ($K^\prime$) are the
valence band maximum (VBM) and the conduction band minimum (CBM),
respectively. The VBM has twofold advantages: In the first there
exists a large SOC-induced valley-contrasting spin splitting of
about 0.41 eV, which is comparable to that of 2\emph{H}-WSe$_2$
monolayer \cite{WSe2_spinlayer_2014,WSe2_NN_2013} in
Fig.~\ref{fig3}(b). In the second, the VBM at $K$ ($K^\prime$) is
0.4 eV higher than that at $\Gamma$ and it is robust against strain
or layer hybridization, which provides a large space for hole doping
to investigate spin-valley physics. Our calculations reveal a Berry
curvature contrasting behavior at $K$ and $K^\prime$ (Inset of
Fig.~\ref{fig3}(a)), which definitely gives rise to the strong
valley Hall effect flowing to opposite transverse edges when an
in-plane electric field is applied
~\cite{xiao_valley-contrasting_2007} and also leads to a stronger
valley-dependent optical selection rule at both $K$ and $K^\prime$
points ~\cite{yao_valley-dependent_2008}. Furthermore, it exhibits a
large hole mobility up to about 460 cm$^2$ V$^{-1}$ s$^{-1}$  and
the electron mobility to about 150 cm$^2$ V$^{-1}$ s$^{-1}$ for both
the armchair and zigzag directions (supplementary Fig. S3 and Table
S4~\cite{SM}). These values are about one and a half times those of
2\emph{H}-WSe$_2$ monolayer \cite{WSe2_electron_2014,
WSe2_comprehensive_2018}.

Similarly, among $MA_2$Z$_4$ monolayer with 32 VEC (\emph{M} =
alkali earth elements (Mg, Ca, Sr) and the IIB-group elements (Zn,
Cd, Hg), \emph{A} = Al and Ga, and \emph{Z} = S, Se, and Te), they
are mostly semiconductor in supplementary Fig. S7 and Table
S3~\cite{SM} also due to the closed-shell electronic configurations
of $M^{2+}$, A$^{3+}$ and Z$^{2-}$. Of course, with increasing
atomic mass their band gaps close to become metallic (see
supplementary Fig. S7~\cite{SM}). During this process, extensive
topological phase transition occurs. For instance, in
Fig.~\ref{topo} (a, b, d, and e), we have compiled the DFT-derived
electronic band structures of two selected monolayer materials of
$\beta_2$-SrGa$_2$S$_4$ and $\beta_2$-SrGa$_2$Se$_4$ without and
with the inclusion of the SOC effect. It can be clearly seen that
the case of $\beta_2$-SrGa$_2$S$_4$ is a direct band gap
semiconductor with a gap of about 0.6 eV and 0.5 eV at $\Gamma$
without and with the SOC inclusion, respectively. However, in the
case of its isoelectronic $\beta_2$-SrGa$_2$Se$_4$ without the SOC
inclusion it is a zero-gap semiconductor due to the degenerate
Se-$p_x$ and $p_y$ orbits exactly crossing the Fermi level. With the
SOC inclusion, its zero gap become open again with a small gap value
of about 68 meV. Importantly, in the case $\beta_2$-SrGa$_2$S$_4$
its CBM at $\Gamma$ comprises with Ga \emph{s}-like orbit and the
VBM consists of degenerate $p_{x,y}$ orbits of S. In contrast, in
$\beta_2$-SrGa$_2$Se$_4$ we have observed an opposite situation at
$\Gamma$ the CBM has the degenerate $p_{x,y}$ orbits of Se, whereas
the VBM now becomes the Ga \emph{s}-like orbit. This fact
demonstrates the occurrence of the electronic band inversion,
implying the possible topological non-trivial feature. Hence, we
have calculated their topological index of $Z_2$ value, indicating
$Z_2$ = 0 for trivial $\beta_2$-SrGa$_2$S$_4$ semiconductor and
$Z_2$ = 1 for non-trivial $\beta_2$-SrGa$_2$Se$_4$. This analysis is
confirmed by the evolution of Wannier charge center shown in Fig
~\ref{topo}(f), thereby indicating that $\beta_2$-SrGa$_2$Se$_4$
monolayer material is a topological insulator. Furthermore, this
monolayer topological insulator of $\beta_2$-SrGa$_2$Se$_4$ has to
exhibit non-trivial topological edge states. As shown in
Fig.~\ref{topo}(c), we have derived the edge states along the
$<$010$>$ boundary using the slab modeling in Fig.~\ref{topo}(g),
which indicates clear topological helical edge states with the
appearance of the Dirac cone. Of course, among all these materials
in supplementary Fig. S7 and Fig. S8~\cite{SM}, some of them can be
attributed to be topological insulators, such as
$\beta_2$-CaGa$_2$Se$_4$ and $\beta_2$-MgGa$_2$Te$_4$ and some are
topological semimetals, such as HgGa$_2$Se$_4$, HgAl$_2$Te$_4$ and
$M$Ga$_2$Te$_4$ (\emph{M} = Mg, Ca, Sr, Zn, Cd, and Hg).

\begin{figure*}
\begin{center}
\includegraphics[width=0.8\textwidth]{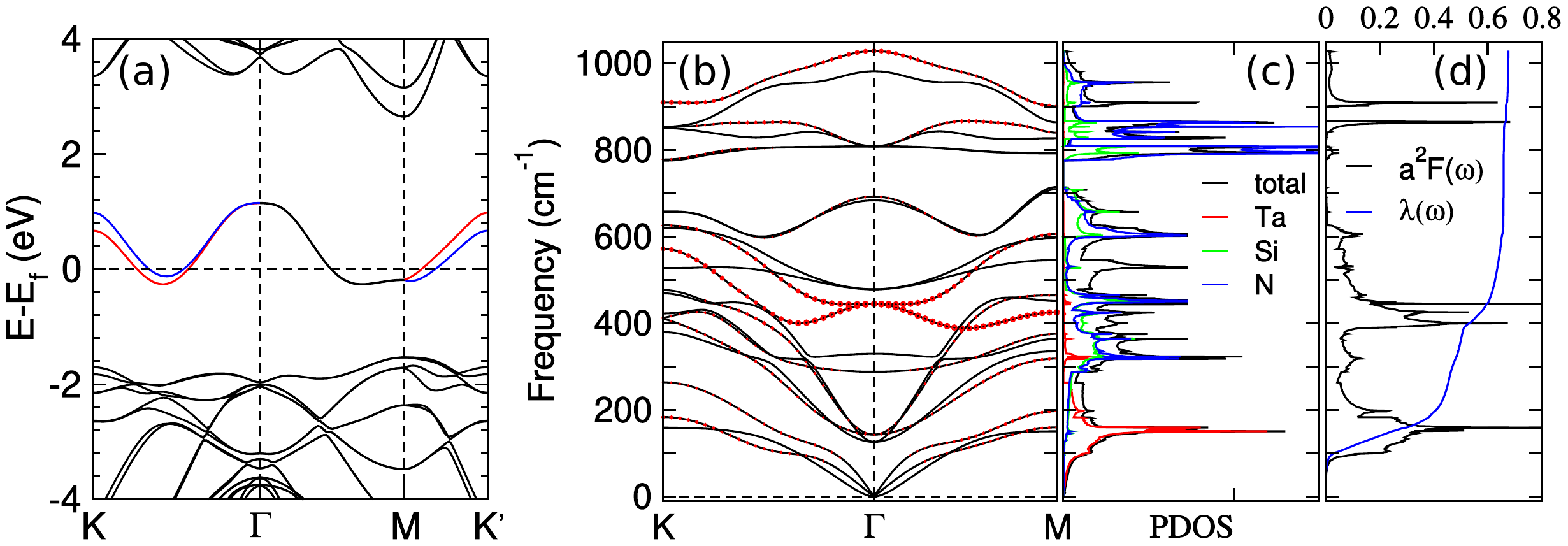}
\caption{(Color online) (a) Contrasting spin-up (red) and spin-down
(blue) split at both $K$ and $K^\prime$ of
$\alpha_1$-TaSi$_{2}$N$_{4}$ due to the SOC-induced Zeeman-like
field, which is opposite for the two valleys. (b) The derived phonon
dispersion in which the area of the red circles represents the
strength of phonon linewidth $\gamma_{\boldsymbol{q}, \nu}$. (c)
Phonon DOS and (d) Eliashberg function $\alpha^{2}F(\omega)$ with accumulated
electron-phonon coupling strength $\lambda(\omega)$.} \label{fig4}
\end{center}
\end{figure*}

\textbf{VEC of 33.} In difference from the systems with 32 or 34
VEC, $MA_2Z_4$ monolayer materials with 33 VEC is very special. This
is mainly because of the existence of one more unpaired electron
than 32 VEC and one less than 34 VEC. In the first, the unpaired
electron has to cross the Fermi level, leading to the metallic
occurrence and, in the second, the one more unpaired electron
provides the crucial prerequisites for the onset of magnetic
ordering. Our spin-polarized calculations reveal that, in the system
of 33 VEC, there are eight ferromagnetic monolayer materials
(supplementary Fig. S5~\cite{SM}). $\beta_2$-VGe$_2$N$_4$ is a
typical half-metallic ferromagnet with an integer spin moment (1.0
$\mu_B$) that V atom carries, because its spin-up band carries
electronic density of states at the Fermi level and its spin-down
band is a semiconductor with a band gap. Both
$\alpha_1$-VSi$_2$N$_4$ and $\alpha_1$-NbGe$_2$N$_4$ seem exactly on
the edge of the half-metallic ferromagnetism and both V and Nb atoms
carry the nearly integer spin moments of 0.97 $\mu_B$ and 0.98
$\mu_B$, respectively. In particular, we need to emphasize that
$\alpha_1$-VSi$_2$P$_4$ is a parabolic spin-gapless ferromagnetic
semiconductor with an total integer spin moment of
1.0$\mu_B$~\cite{wang_proposal_2008,Wang2017} in
Fig.~\ref{fig3}(c,d,e). Although both majority and minority channels
are semiconductor, the VBM of the majority electrons touches the
Fermi level at the $K$ or $K^\prime$ points and the CBM of the
minority electrons touches the Fermi level as well, but at the M
point. This fact means that for an excitation energy up to the band
gap energy of the other spin channel, the excited electrons and
holes are both 100\% spin
polarized~\cite{wang_proposal_2008,Wang2017}. Interestingly, if we
continuously check $MA_2Z_4$ monolayer materials with its VEC from
32 to 38 in this series of $M$ = Ti, V, Cr, Mn, Fe, Co, and Ni, we
have found more half-metallic ferromagnetism satisfying the
Slater-Pauling behavior that allows us to estimate the total spin
moment via $M_t$ = 3-$|$VEC-35$|$ $\mu_B$.

Nonmagnetic metallic $MA_2Z_4$ with 33 VEC are intriguing. Within
this kind of nonmagnetic situation of the noncentrosymmetric 2D
lattice, an unpaired electron will result in a half-filled
electronic band which crosses the Fermi level. Similar to
$\alpha_2$-WSi$_{2}$P$_{4}$ monolayer semiconductor in
Fig.~\ref{fig3}(a), the SOC effect of Ta atom also induces a very
large valley-contrasting spin splitting at the $K$ and $K$$^\prime$,
contributing Zeeman-like spin splittings (Fig.~\ref{fig4}(a)). This
will be beneficial to the occurrence of Ising superconductivity, as
what one already observed in TMDC
NbSe$_2$\cite{Type-II-PRL-wang,Lu1353,
Ising-Zhou-PRB,saito_NPs_2016,Inte-Liu-PRX,xi_ising_2016,
de_tuning_2018,falson_type-ii_2020}. Following this inspiration, we
have further derived the phonon dispersion
and Eliashberg function ($\alpha^2$F($\omega$)) as well as
accumulated electron-phonon coupling strength ($\lambda$) in
Fig.~\ref{fig4}(b to c) for the nonmagnetic metallic
$\alpha_1$-TaSi$_2$N$_4$. Using the total $\lambda$=0.68 and the
calculated logarithmic average phonon frequency of 298.8 $cm^{-1}$,
we have derived the superconductive transition temperature $T_c$ =
9.67 K via the Dynes modified McMillan formula with the effective
screened Coulomb repulsion constant of $\mu$ =0.10. The
superconductive $\alpha_1$-TaSi$_2$N$_4$ monolayer material is
remarkable, because in its noncentrosymmetric lattice with a large
SOC splitting spins of Cooper pairs are aligned along the
out-of-plane direction in accompanying with a large in-plane upper
critical field exceeding the Pauli paramagnetic limit. Its
superconductivity effectively does not respond to the in-plane
magnetic field, which make its superconductivity robust against a
magnetic filed \cite{Type-II-PRL-wang,Lu1353,
Ising-Zhou-PRB,saito_NPs_2016,Inte-Liu-PRX,xi_ising_2016,
de_tuning_2018,falson_type-ii_2020}. Of course, our calculations
still reveal that the other two $\alpha_2$-TaGe$_2$P$_4$ and
$\beta_2$-HfGe$_2$P$_4$ are superconductor with the estimated $T_c$
of 3.75 K and 1.07 K, as show in supplementary Fig. S6~\cite{SM},
respectively.

\section{Conclusion}
In this work, we present a general intercalated architecture
approach to construct septuple-atomic-layer MA$_2$Z$_4$ monolayer
materials. Our approach predicts 66 $MA_2Z_4$ monolayer materials
which are stable both thermodynamically and dynamically, among 72
candidates considered here. Interestingly, their electronic
properties spans a wide range upon the valence electron number per
formula (VEC). $MA_2Z_4$ monolayer materials with the rules of 32 or
34 VEC are mostly semiconductors with direct or indirect band gap.
Upon the spin-orbit coupling strength associated with the atomic
mass, topological transitions have been predicted to occur from
trivial semiconductors to non-trivial topological insulators,
\emph{e.g.}, from trivial semiconductor of $\beta_2$-SrGa$_2$S$_4$
to non-trivial topological insulator of $\beta_2$-SrGa$_2$Se$_4$ to
topological semimetal of $\beta_2$-SrGa$_2$Te$_4$. In addition,
these 2D semiconductors with noncentrosymmetric in-planar lattices
provide plenty of room to study spin-valley coupling physics due to
the momentum-contrasting spin-valley splitting and Berry curvatures
at $K$ or $K^\prime$ point of the 2D hexagonal BZ corners
(\emph{e.g.}, $\alpha_1$-WSi$_2$P$_4$). We also predict that
$MA_2Z_4$ monolayer materials with 33 VEC are general metal, or
half-metal ferromagnetism (\emph{e.g.}, $\alpha_1$-VSi$_2$N$_4$), or
spin-gapless semiconductor (\emph{e.g.}, $\alpha_1$-VSi$_2$P$_4$)
upon whether or not an unpaired electron is spin polarized.
Significantly, our calculations even suggest the existence of the
intrinsic Ising superconductor in metallic $\alpha_1$-TaSi$_2$N$_4$
monolayer material. It is mainly because that superconductive cooper
pairs formed from carriers in intrinsic spin-orbit coupling valleys
at \emph{K} and \emph{K}$^\prime$ points exhibit locked opposite
spins. This behavior implies that its superconductivity has no
responsibility to an in-plane pair-breaking field such as magnetic
field, which can remarkably enhance its in-plane upper critical
field.

Finally, we would like to emphasize that our currently proposed
intercalated architecture approach can be indeed extended to
$MA_2Z_4$ monolayer materials with \emph{M} for late transition
metal elements, such as MnBi$_2$Te$_4$ for which our current
calculations also correctly capture the agreements to experiments.
Furthermore, it can be further generalized to a wider way. For
instance, $n$=7 $MA_2Z_4$ monolayer materials can be also
constructed by intercalating \emph{n} = 2 germanene monolayer into
\emph{n}=5 vdW Bi$_2$Se$_3$ monolayer structure and, we can even
combine \emph{n}=3 MoS$_2$ monolayer materials and \emph{n}=5 vdW
Bi$_2$Se$_3$ monolayer structure to form new type of \emph{n}=8
monolayer materials, and so on.

\section{Method}
\textbf{ Electronic and phononic band structures.}
First-principles calculations were employed using the Vienna ab \emph{initio}
simulation package (VASP) \cite{kresse-PRB-1996,kresse-PRB-1999}
with the exchange-correlation (XC) potential of Perdew-Burke-Ernzerhof
(PBE) type  and projector augmented wave (PAW) method. Since the transition
metal element $M$ in $MA_2Z_4$ monolayer have a larger atomic mass, we also
considered the spin-orbit coupling (SOC) to calculate electronic band
structure. Furthermore,
in order to get more exact bandgap, hybrid Heyd-Scuseria-Ernzerhorf (HSE06)
functionals are also employed. Taken all the elements into consideration,
the 500 eV cutoff energy was chose. And the $k$-point sampling grid in the
self-consistent process was 15 $\times$ 15 $\times$ 1 in $\Gamma$-centered
Monkhorst-Pack scheme. The Force convergence criteria on each atom is less than
10$^{-3}$ eV/{\AA} and the energy convergence criteria on the primitive cell is less
10$^{-6}$ eV. To minimize the interactions between the layer with its
periodic images, a vacuum of 20 {\AA} between layers was considered.

Phonon spectra was obtained using the density
functional perturbation theory (DFPT) method implemented in Phonopy
\cite{phonopy_2015} package. 4 $\times$ 4 $\times$ 1 and 5 $\times$
5 $\times$ 1 supercell are used for the calculation of the phonon
spectra to make sure that the force constants are sufficiently collected.
In addition, we applied an iterative Green functions method \cite{GreenF}to
calculate the edge states and used the Wannier charge centers method
introduced in Ref.\citenum{WCC} to obtain the $Z_2$ value.

\textbf{Electron-phonon coupling and superconductivity.}
For metallic materials, the electron-phonon coupling (EPC) constant
$\lambda(\omega)$ \cite{Eliashberg1960} is given by
\begin{equation}\label{lambda}
\lambda(\omega)= 2 \int d \omega \alpha^{2} F(\omega) / \omega
\end{equation}
where $\alpha^{2} F(\omega)$ is the Eliashberg function and defined as

\begin{equation}\label{a2F}
\alpha^{2} F(\omega)=\frac{1}{2 \pi N\left(\epsilon_{\mathrm{F}}\right)} \sum_{\boldsymbol{q} \nu} \delta\left(\omega-\omega_{\boldsymbol{q} \nu}\right) \frac{\gamma_{\boldsymbol{q} \nu}}{\hbar \omega_{\boldsymbol{q} \nu}}
\end{equation}

where $N(\epsilon_{F})$ is density of states (DOS) at Fermi level, $\omega_{\boldsymbol{q} \nu}$ is
phonon frequency of the mode $\nu$ at the wavevector $\boldsymbol{q}$ and
$\gamma_{\boldsymbol{q} \nu}$ is phonon linewidth or lifetime.

Based on BCS theory, the results of Eliashberg function $\alpha^{2} F(\omega)$ can be
used to calculate logarithmic average phonon frequencies by $\omega_{\log } = \exp \left[\frac{2}{\lambda}
\int_{0}^{\infty} \frac{\mathrm{d} \omega}{\omega} \alpha^{2} F(\omega)\log \omega\right]$
and, further, to calculate the superconductivity critical temperature, $T_{\mathrm{c}}
=\frac{\omega_{\log }}{1.2} \exp \left[\frac{-1.04(1+\lambda)}{\lambda-\mu^{*}
(1+0.62 \lambda)}\right]$, by using the simple Allen-Dynes-modified McMillan
formula \cite{allen_Tc_1975}.

The EPC in this work calculated with local density
approximation \cite{LDA-1981} as
implemented in the Quantum-ESPRESSO \cite{QE-2009} package with
Norm-conserving pseudopotentials (NCPP). For $\alpha_1$-TaSi$_2$N$_4$ and
$\beta_2$-HfGe$_2$P$_4$, the kinetic energy cutoff and the charge density cutoff
of the plane wave basis are chosen to be 60 and 480 Ry. 32 $\times$ 32 $\times$ 1
k-mesh with Marzari-Vanderbilt cold smearing of 0.02 Ry is used
to evaluate the self-consistent electron density. 4 $\times$ 4 $\times$ 1 q-mesh
are used to obtain the dynamic matrix and EPC constant, respectively. For
$\alpha_2$-TaGe$_2$P$_4$, due to the softening of its acoustic mode, 80 Ry kinetic
energy cutoff, 640 Ry charge density cutoff, 36 $\times$ 36 $\times$ 1 k-mesh,
6 $\times$ 6 $\times$ 1 q-mesh are used to calculate its EPC strength and its $T_c$.

\textbf{Carrier mobilities.}
Intrinsic carrier mobility $\mu$ of 2D materials based on deformation potential
is calculated by \cite{bardeen_deformation_1950}
\begin{equation}\label{carriermobility}
\mu_{2 D}=\frac{2 e \hbar^{3} C}{3 k_{\mathrm{B}} T\left|m^{*}\right|^{2} E_{1}^{2}}
\end{equation}
where $C$ is the elastic modulus defined as $\left[\partial^{2} E / \partial \delta^{2}\right] / S_{0}$ ,
$m^{*}$ is the effective mass at conduction band minimum (CBM) or
valance band maximum (VBM), and $T$ is the temperature, here
room temperature $T$ = 300 K was used. $E_{1}$ is the deformation potential (DP)
constant defined as $\Delta E /\left(\Delta l / l_{0}\right)$ , where $\Delta E$
is the change of the eigenvalue at CBM or VBM and $\Delta l$ is the lattice dilation
along deformation direction.

\textbf{Enthalpies of formation.}
Enthalpies of formation ($E_{f}$) per atom can be expressed by the following equation:
\begin{equation}\label{Ecohesive}
E_{f} = \{E_{\mathrm{tot}}-\left(E_{\mathrm{M}} + 2E_{\mathrm{A}}+ 4E_{\mathrm{Z}}\right)\} /7
\end{equation}
Where $E_{tot}$ is the total energy of the system, and $E_{M}$, $E_{A}$ and $E_{Z}$
are the ground state energies of elementary substances $M$, $A$ and $Z$ in $MA_{2}X_{4}$
compounds, respectively.

\textbf{Berry curvature calculation.}
The Berry curvature of a 2D material with $n$ bands can be defined as
\cite{xiao_berry_2010,tunable_MoS2_2018}:
\begin{equation}\label{BerryCurvature}
\Omega_{z}(\mathbf{k})=\nabla_{\mathbf{k}} \times i\left\langle u_{n, \mathbf{k}} |
\nabla_{\mathbf{k}} u_{n, \mathbf{k}}\right\rangle
\end{equation}
where $u_{n, \mathbf{k}}$ is the lattice periodic part of the Bloch wave functions.
And it can be derived by a tight binding Hamiltonian obtained
from first principle calculations via maximally-localized Wannier
functions method\cite{MLWF-1997}. Here, we used all occupied bands
 and the approach introduced by Ref. \citenum{wang_ab_2006}
to calculate the Berry curvature.

\section{Acknowledgments}
Work was supported by the National Science Fund for Distinguished
Young Scholars (grant number 51725103), by the National Natural
Science Foundation of China (grant number 51671193), by the Science
Challenging Project (grant number TZ2016004), and by major research
project 2018ZX06002004. M.C. was supported by the National Natural
Science Foundation of China (Grant No. 11774084) and the Project of
Educational Commission of Hunan Province of China, 18A003. All
calculations have been performed on the high-performance
computational cluster in the Shenyang National University Science
and Technology Park.

\section{Author contributions}
X.-Q.C. supervised this project. Structural model and structural
optimization as well as band structure calculations were performed by
L.W.. Y.S. and Q.G. carried out theoretical calculations of carrier
mobility and M.L. and R.L. performed detailed calculations of
electron-phonon coupling. L.W. and X.-Q. C. wrote the manuscript
with contributions from all authors. All authors including Y.-L.H,
W.R. ,M.C. ,H.-M.C. and Y.L. discussed the results and commented on the
manuscript.
\section{Declaration of interests}
The authors declare no competing financial interests.

\clearpage

\setlength{\baselineskip}{20pt}

{\large \textbf{Supplementary Materials including as follows,}}


{\large \textbf{1) Supplementary Figures S1 (page 11) to S8 (page 17)}}

{\large \textbf{2) Supplementary Tables S1 (page 19) to S4 (page 22)}}



\begin{figure*}
\begin{center}
\includegraphics[width=0.98\textwidth]{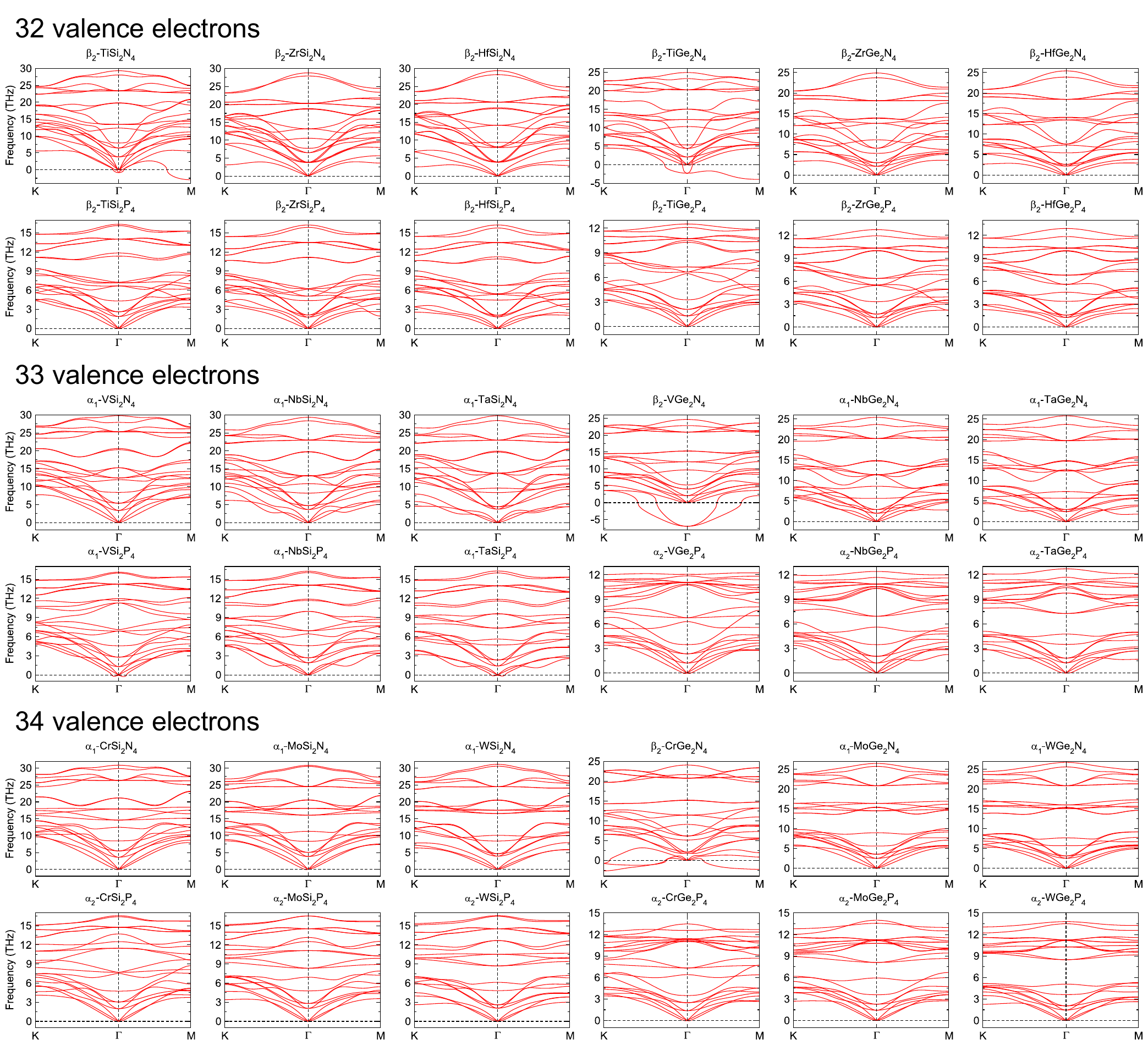}
\parbox[c]{17.0cm}{{\bf Fig. S1,} 
The phonon spectrum of 36 septuple-atomic-layer $MA_2Z_4$ monolayers listed in \textbf{\textcolor{blue}{Table S2}}.}
\label{phonons} 
\end{center}
\end{figure*}

\newpage
\textcolor{white}{123}
\begin{figure*}
\begin{center}
\includegraphics[width=1.0\textwidth]{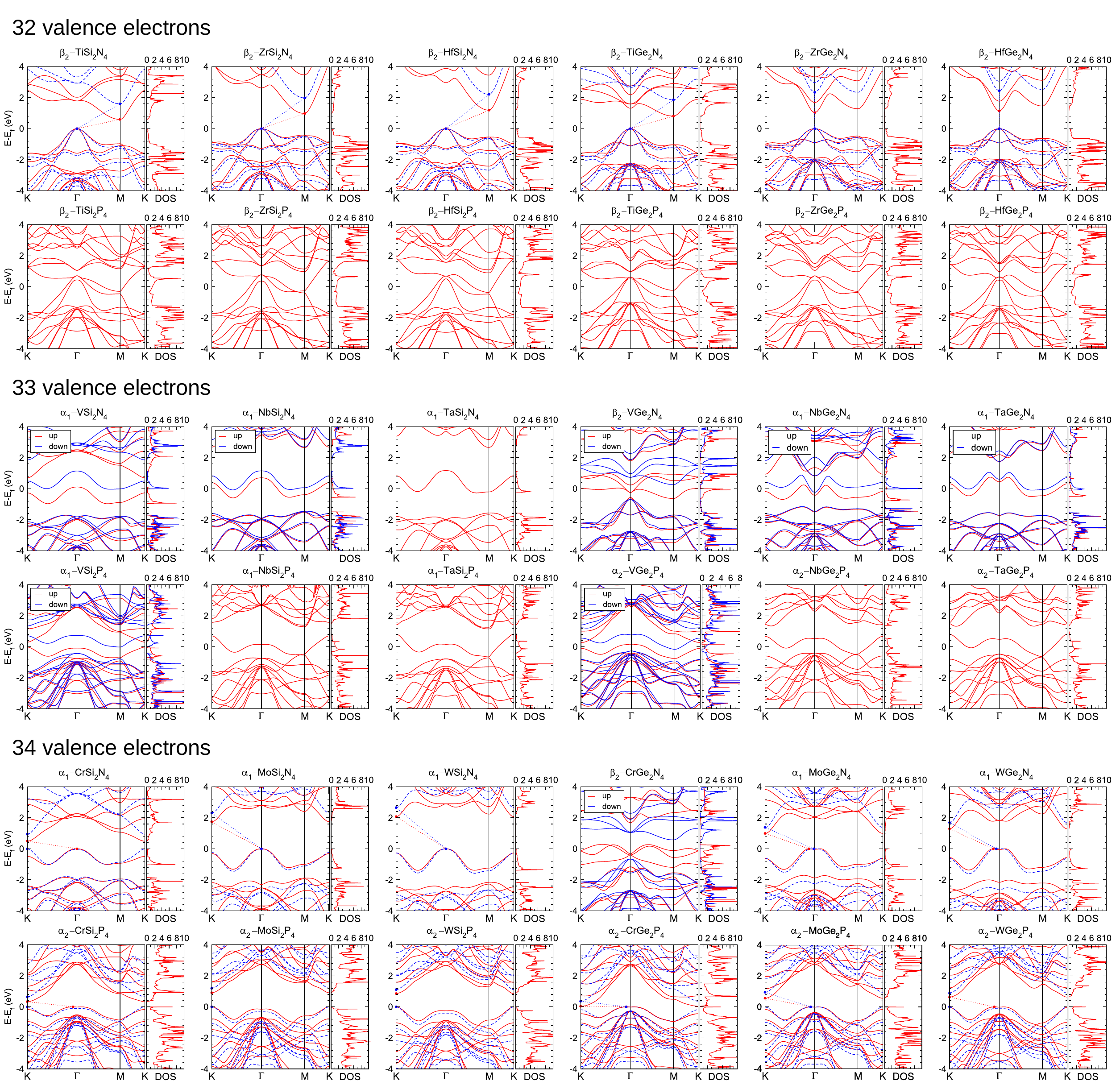}
\parbox[c]{17.0cm}{{\bf Fig. S2,} \normalsize(color online) The electronic structures
of 36 septuple-layer $MA_2Z_4$ monolayers listed in \textbf{\textcolor{blue}{Table S2}},
where, for semiconductors, the red solid and blue dash bands are calculated by PBE and
HSE06 functional without inclusion of spin orbit coupling (SOC) and, for ferromagnetic metallic compounds,
the red and blue solid bands represent spin up and spin down bands. }
\label{bands}
\end{center}
\end{figure*}

\begin{figure*}
\begin{center}
\includegraphics[width=0.85\textwidth]{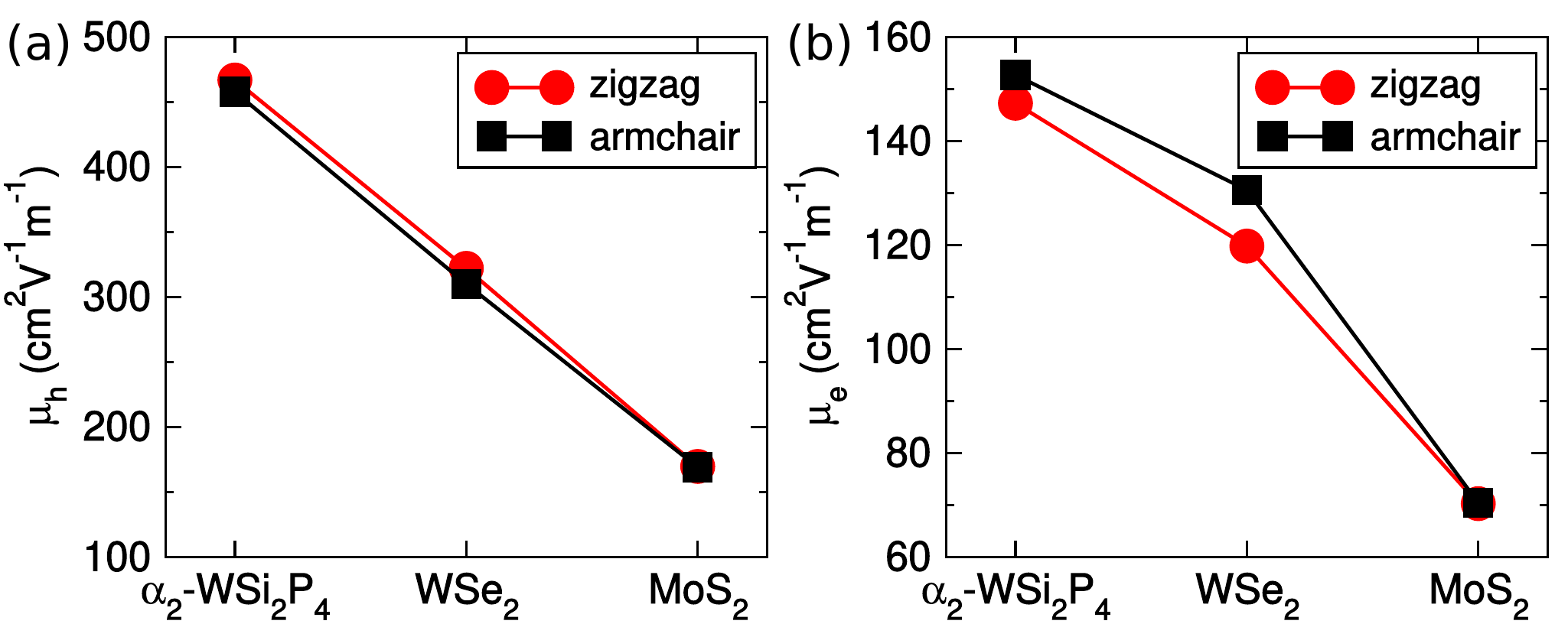}
\parbox[c]{17.0cm}{{\bf Fig. S3,} \normalsize(color online) Hole and electron
mobility of $\alpha_2$-WSi$_2$P$_4$, WSe$_2$ and MoS$_2$. Note that all the detail
data used to calculate hole and electron mobility are listed in \textbf{\textcolor{blue}{Table S4} }}
\label{dos-mag}
\end{center}
\end{figure*}

\begin{figure}
\begin{center}
\includegraphics[width=0.85\textwidth]{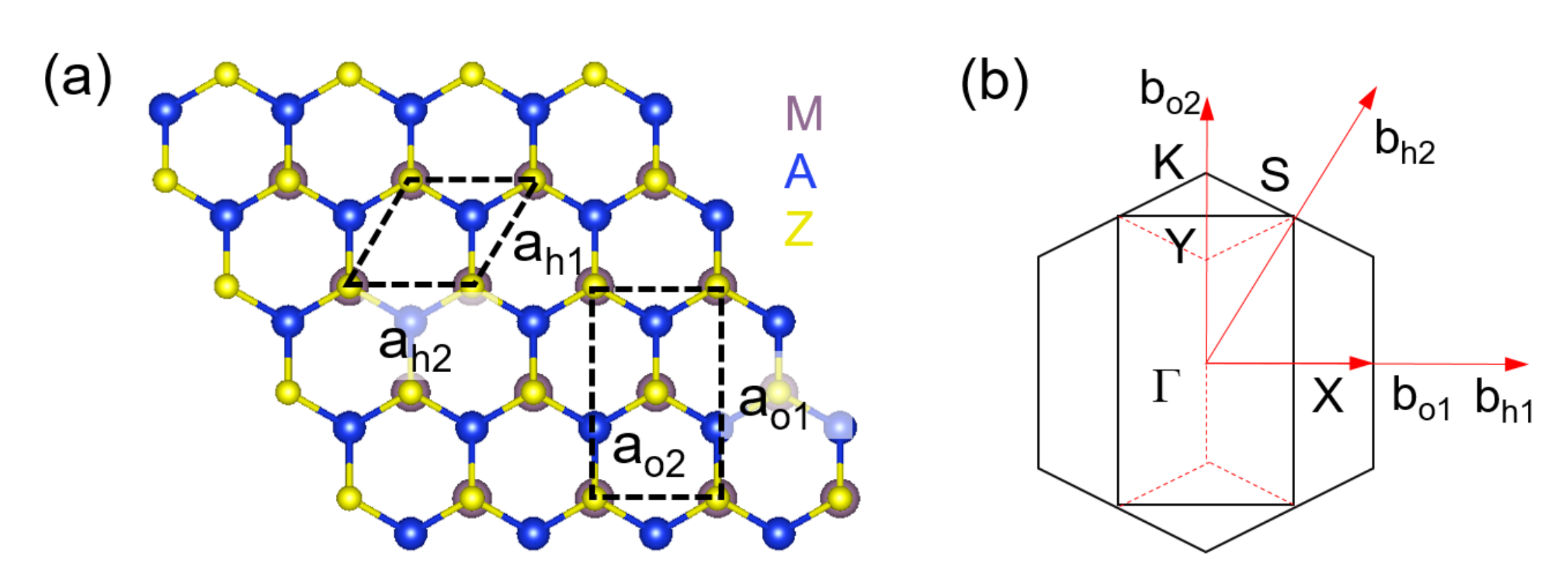}
\parbox[c]{17.0cm}{{\bf Fig. S4,} \normalsize(color online) (a) The top view of $\alpha_2$-$MA_{2}Z_{4}$
monolayer. The dashed black lines represent the hexagonal primitive cell and orthogonal
supercell. (b) The FBZ of these two lattices. The dashed red line shows the folding of FBZ of
hexagonal cell into FBZ of orthogonal cell.}
\label{cmm}
\end{center}
\end{figure}

\begin{figure*}
\begin{center}
\includegraphics[width=0.85\textwidth]{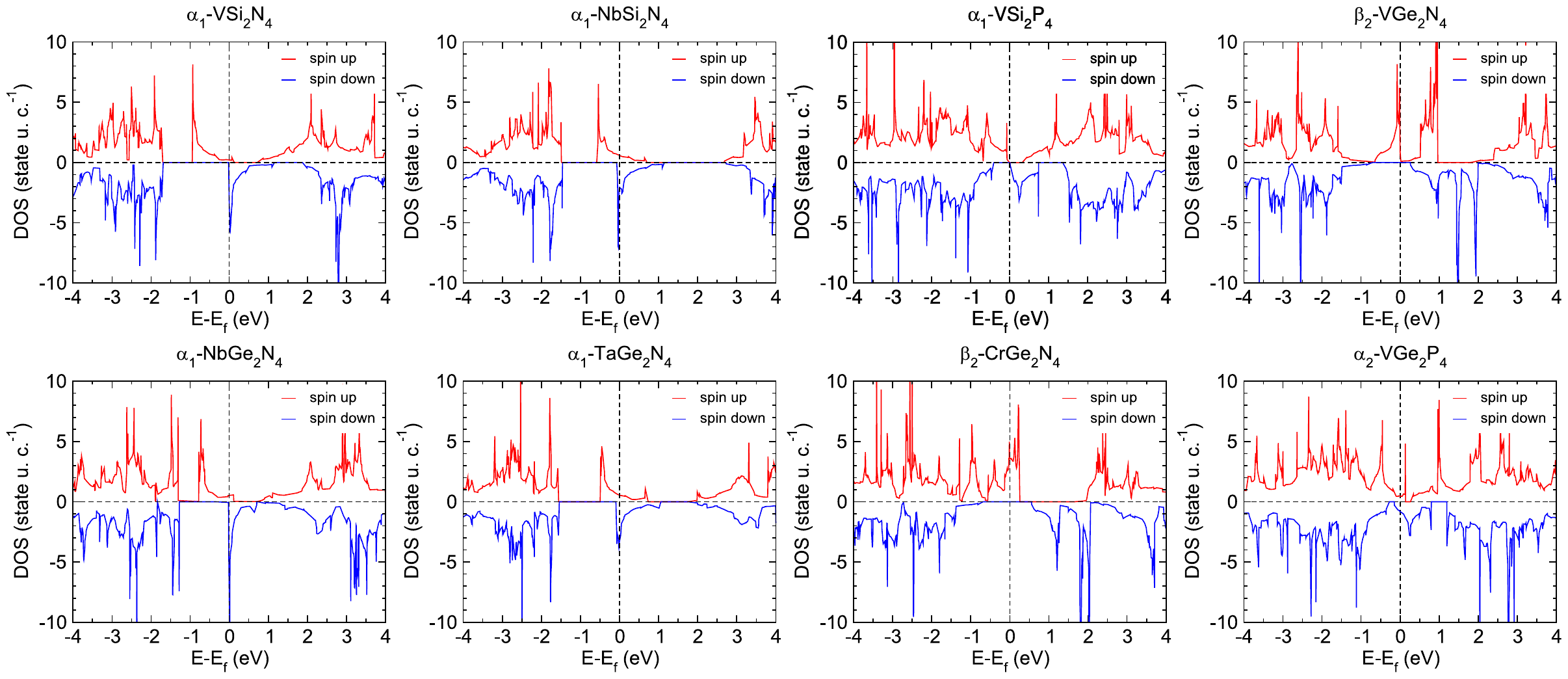}
\parbox[c]{17.0cm}{{\bf Fig. S5,} \normalsize(color online) The DOS of eight ferromagnetic $MA_2Z_4$ monolayers.}
\label{dos-mag}
\end{center}
\end{figure*}

\begin{figure*}
\begin{center}
\includegraphics[width=0.85\textwidth]{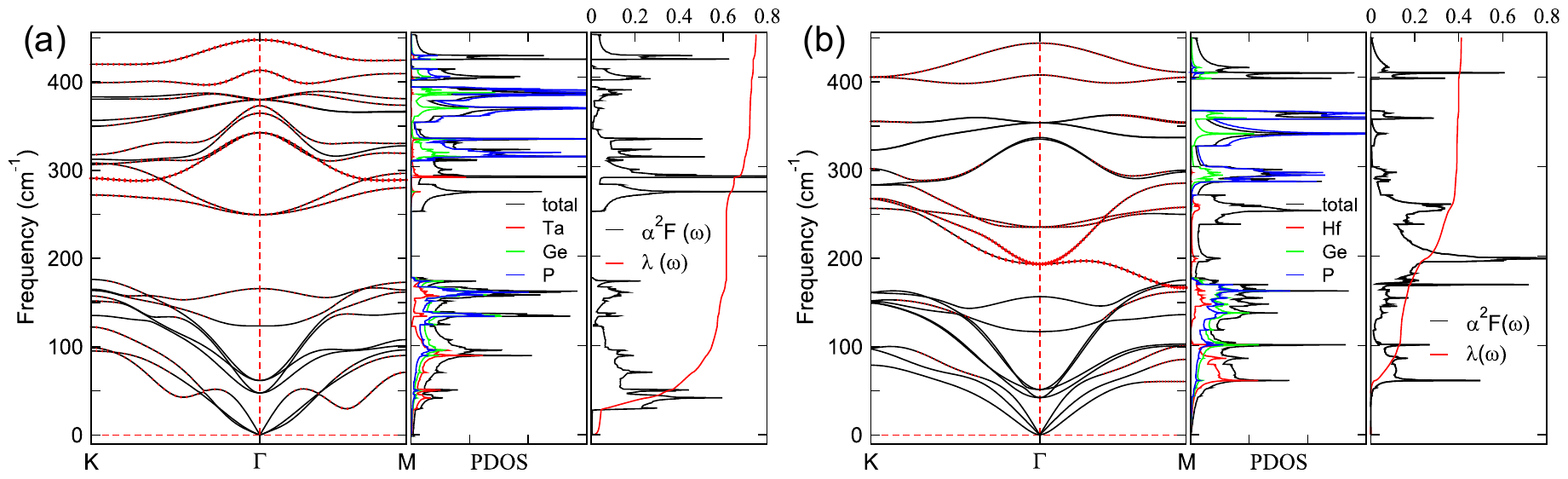}
\parbox[c]{18.0cm}{{\bf Fig. S6,} \normalsize(color online) The phonon dispersion,
phonon DOS and Eliashberg function $\alpha^{2}F(\omega)$ with EPC strength $\lambda(\omega)$
of the $\alpha_2$-TaGe$_2$P$_4$ (a) and $\beta_2$ HfGe$_{2}$P$_{4}$ (b), where
the area of the red circles represents the strength of phonon linewidth
$\gamma_{\boldsymbol{q}, \nu}$,}
\label{epc}
\end{center}
\end{figure*}

\textcolor{white}{123}
\begin{figure*}
\begin{center}
\includegraphics[width=0.95\textwidth]{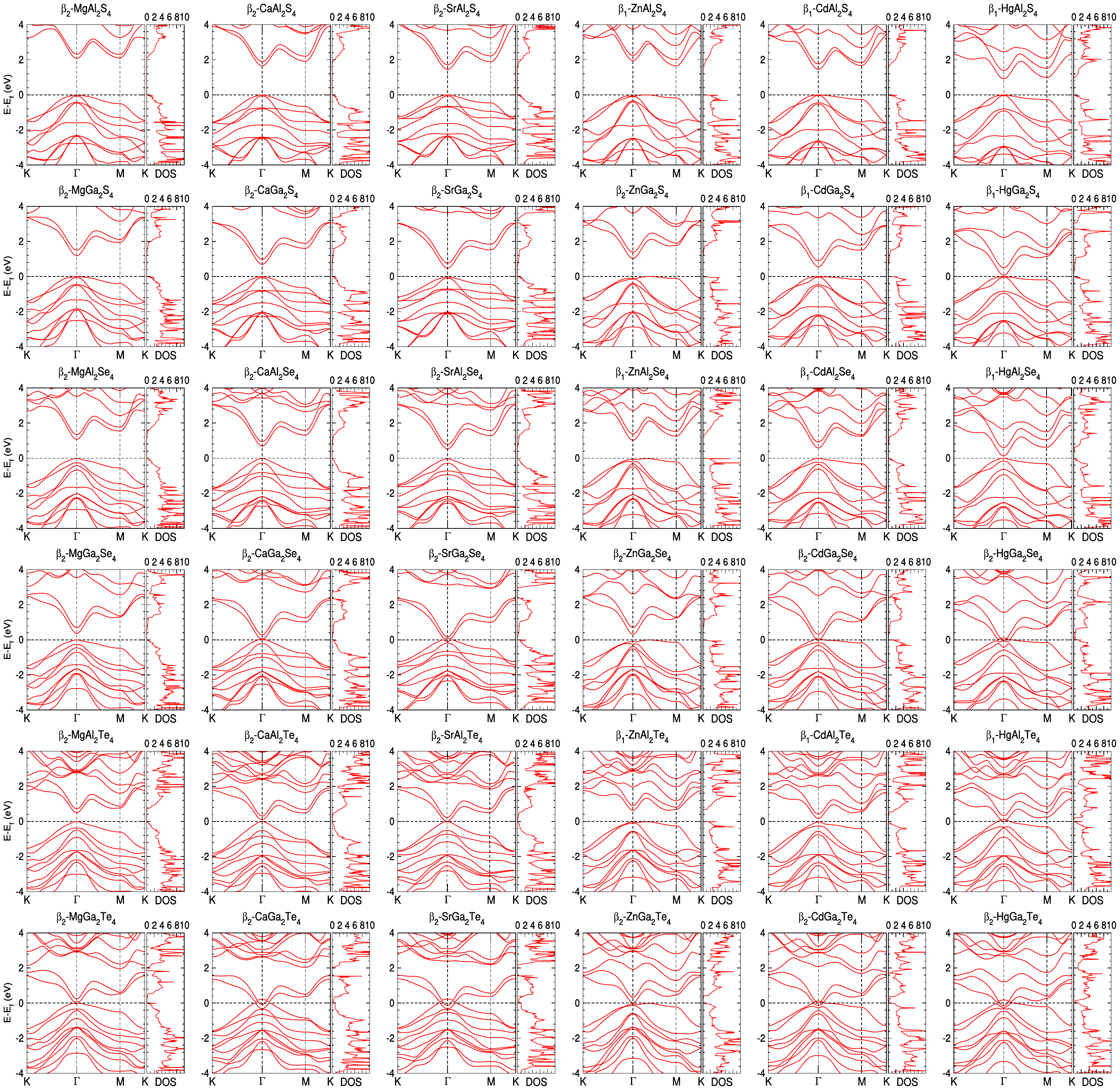}
\parbox[c]{17.0cm}{{\bf Fig. S7,} \normalsize(color online) The electronic structures of
36 energetically favorable septuple-layer $MA_2Z_4$ monolayers, as listed in \textbf{\textcolor{blue}{Table S3}},
with inclusion of spin orbit coupling.}
\label{phonons}
\end{center}
\end{figure*}

\textcolor{white}{123}
\begin{figure*}
\begin{center}
\includegraphics[width=0.85\textwidth]{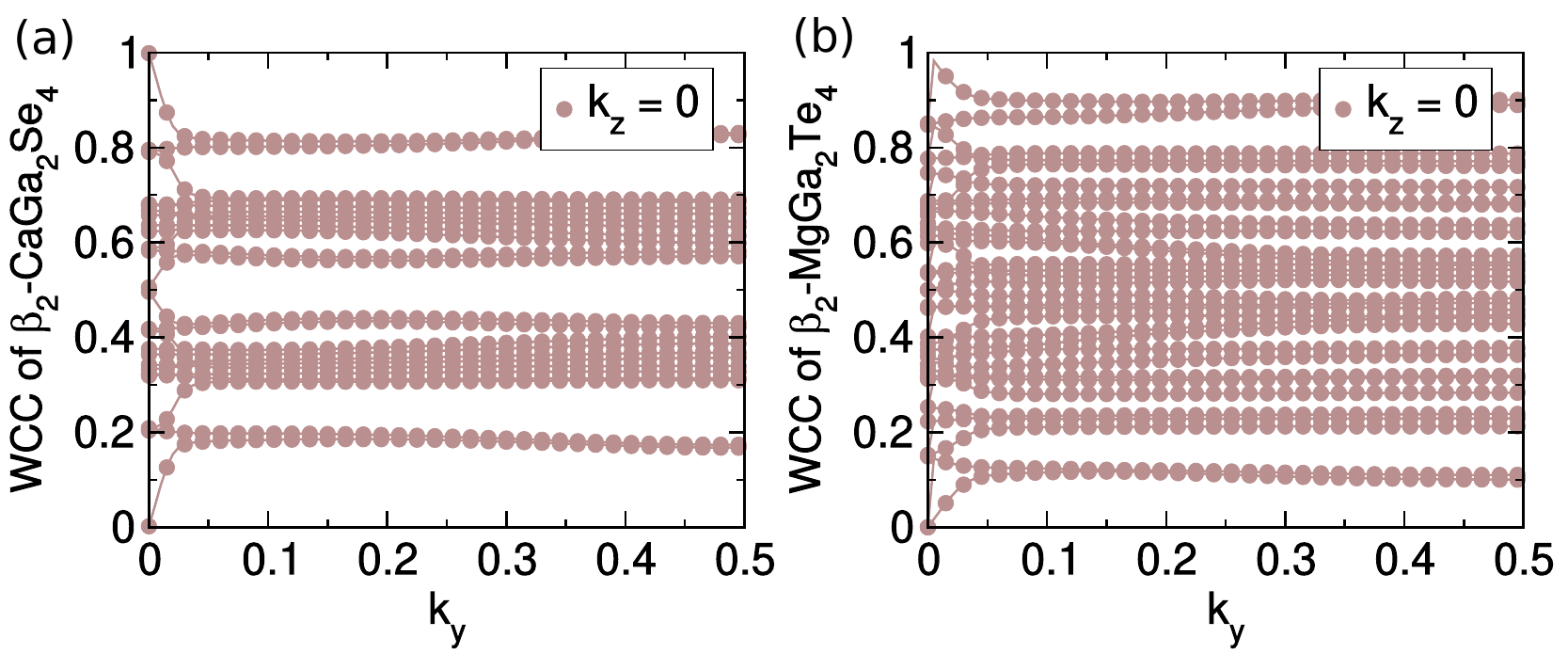}
\parbox[c]{17.0cm}{{\bf Fig. S8,} \normalsize(color online) Evolution of Wannier charge center (WCC) of
$\beta_2$-CaGa$_2$Se$_4$ (a) and $\beta_2$-MgGa$_2$Te$_4$ (b) in the k$_z$ = 0 plane, which implies a
nonzero topological invariant.}
\label{phonons}
\end{center}
\end{figure*}

\clearpage

\begin{table*}[htp]
\setlength{\tabcolsep}{1.5mm}
\begin{center}
\parbox[c]{18.0cm}{{\bf Table S1,} \normalsize The enthalpies of formation E$_f$ (eV/atom) of $\alpha_{1-6}$-
and $\beta_{1-6}$-$MA_2Z_4$ monolayers with space group P\={6}m2 and P\={3}m1, respectively, where $M$ = Ti, Zr, Hf, V, Nb, Ta, Cr, Mo, W; $A$ = Si and Ge; $Z$ = N and P.} \label{formation-Mo}
\begin{tabular}{clcccccccccccc}
\hline\hline
No.&  Name & $\alpha_1$ & $\alpha_2$ & $\alpha_3$ & $\alpha_4$ & $\alpha_5$ & $\alpha_6$ & $\beta_1$& $\beta_2$& $\beta_3$& $\beta_4$& $\beta_5$& $\beta_6$ \\\hline
01& TiSi$_2$N$_4$ &-1.052 &-1.036 &-0.262 &-0.633 &-0.668 &-0.282 &-1.091 &\textbf{-1.119} &-0.765 &-0.405 &-0.708 &-0.369 \\
02& ZrSi$_2$N$_4$ &-0.991 &-0.986 &-0.183 &-0.558 &-0.570 &-0.196 &-1.013 &\textbf{-1.028} &-0.631 &-0.272 &-0.595 &-0.235 \\
03& HfSi$_2$N$_4$ &-1.053 &-1.047 &-0.217 &-0.602 &-0.617 &-0.231 &-1.091 &\textbf{-1.105} &-0.694 &-0.333 &-0.661 &-0.298 \\
04& TiGe$_2$N$_4$ &-0.331 &-0.326 &0.242  &-0.009 &-0.021 &0.087  &-0.395 &\textbf{-0.409} &-0.142 &0.083  &-0.105 &0.115\\
05& ZrGe$_2$N$_4$ &-0.378 &-0.376 &0.220  &-0.040 &-0.039 &0.221  &-0.426 &\textbf{-0.433 }&-0.133 &0.098  &-0.113 &0.128\\
06& HfGe$_2$N$_4$ &-0.425 &-0.423 &0.193  &-0.073 &-0.074 &0.195  &-0.489 &\textbf{-0.496 }&-0.183 &0.051  &-0.165 &0.077\\
07& TiSi$_2$P$_4$ &-0.327 &-0.316 &0.171  &-0.034 &-0.204 &-0.101 &-0.335 &\textbf{-0.348 }&-0.193 &-0.010 &-0.073 &0.075\\
08& ZrSi$_2$P$_4$ &-0.366 &-0.349 &0.080  &-0.077 &-0.206 &-0.129 &-0.372 &\textbf{-0.390 }&-0.252 &-0.074 &-0.114 &0.043\\
09& HfSi$_2$P$_4$ &-0.357 &-0.342 &0.122  &-0.060 &-0.155 &-0.095 &-0.366 &\textbf{-0.383 }&-0.231 &-0.048 &-0.104 &0.049\\
10& TiGe$_2$P$_4$ &-0.194 &-0.190 &0.153  &0.020  &-0.117 &-0.018 &-0.196 &\textbf{-0.200 }&-0.011 &0.004  &0.051  &0.123\\
11& ZrGe$_2$P$_4$ &-0.253 &-0.244 &0.023  &0.017  &-0.127 &-0.132 &-0.253 &\textbf{-0.262 }&-0.096 &-0.096 &-0.013 &0.077\\
12& HfGe$_2$P$_4$ &-0.243 &-0.235 &0.066  &0.026  &-0.106 &-0.091 &-0.249 &\textbf{-0.258 }&-0.076 &-0.070 &-0.006 &0.083\\
13& VSi$_2$N$_4$  &\textbf{-0.954 }&-0.933 &-0.196 &-0.545 &-0.619 &-0.231 &-0.913 &-0.949 &-0.637 &-0.252 &-0.540 &-0.211 \\
14& NbSi$_2$N$_4$ &\textbf{-0.989 }&-0.964 &-0.206 &-0.552 &-0.628 &-0.245 &-0.927 &-0.966 &-0.629 &-0.243 &-0.534 &-0.206 \\
15& TaSi$_2$N$_4$ &\textbf{-1.009 }&-0.978 &-0.205 &-0.551 &-0.634 &-0.244 &-0.944 &-0.984 &-0.633 &-0.252 &-0.537 &-0.222\\
16& VGe$_2$N$_4$  &-0.173 &-0.167 &0.359  &0.138  &0.098  &0.320  &-0.171 &\textbf{-0.186 }&0.057  &0.280  &0.123  &0.312\\
17& NbGe$_2$N$_4$ &\textbf{-0.293 }&-0.280 &0.254  &0.036  &-0.016 &0.222  &-0.255 &-0.275 &-0.029 &0.173  &0.042  &0.186\\
18& TaGe$_2$N$_4$ &\textbf{-0.310 }&-0.295 &0.238  &0.031  &-0.027 &0.214  &-0.276 &-0.298 &-0.043 &0.134  &0.028  &0.144\\
19& VSi$_2$P$_4$  &\textbf{-0.246 }&-0.244 &0.112  &0.026  &-0.111 &0.023  &-0.221 &-0.225 &-0.061 &0.127  &0.009  &0.157\\
20& NbSi$_2$P$_4$ &\textbf{-0.321 }&-0.317 &0.210  &-0.049 &-0.156 &0.005  &-0.290 &-0.294 &-0.138 &0.087  &-0.065 &0.081\\
21& TaSi$_2$P$_4$ &\textbf{-0.297 }&-0.293 &0.256  &-0.025 &-0.113 &0.050  &-0.270 &-0.270 &-0.101 &0.123  &-0.048 &0.095\\
22& VGe$_2$P$_4$  &-0.091 &\textbf{-0.098 }&0.130  &0.117  &0.011  &0.059  &-0.069 &-0.063 &0.114  &0.142  &0.145  &0.228\\
23& NbGe$_2$P$_4$ &-0.176 &\textbf{-0.183 }&0.211  &0.060  &-0.028 &0.019  &-0.151 &-0.139 &0.053  &0.105  &0.053  &0.142\\
24& TaGe$_2$P$_4$ &-0.153 &\textbf{-0.161 }&0.258  &0.081  &0.017  &0.069  &-0.131 &-0.122 &0.078  &0.140  &0.064  &0.153\\
25& CrSi$_2$N$_4$ &\textbf{-0.831} &-0.807 &-0.082 &-0.429 &-0.512 &-0.119 &-0.731 &-0.762 &-0.458 &-0.037 &0.106  &-0.014\\
26& MoSi$_2$N$_4$ &\textbf{-0.955 }&-0.931 &-0.157 &-0.517 &-0.591 &-0.189 &-0.775 &-0.803 &-0.470 &-0.057 &-0.388 &-0.052\\
27& WSi$_2$N$_4$  &\textbf{-0.955 }&-0.929 &-0.147 &-0.502 &-0.579 &-0.177 &-0.757 &-0.782 &-0.434 &-0.033 &-0.363 &-0.042\\
28& CrGe$_2$N$_4$ &0.004  &0.009  &0.514  &0.299  &0.242  &0.479  &-0.001 &\textbf{-0.014 }&0.263  &0.479  &0.308  &0.505\\
29& MoGe$_2$N$_4$ &\textbf{-0.185 }&-0.177 &0.355  &0.136  &0.082  &0.327  &-0.044 &-0.056 &0.187  &0.410  &0.249  &0.427\\
30& WGe$_2$N$_4$  &\textbf{-0.187 }&-0.178 &0.352  &0.144  &0.088  &0.327  &-0.030 &-0.042 &0.205  &0.393  &0.266  &0.408\\
31& CrSi$_2$P$_4$ &-0.176 &\textbf{-0.184 }&0.186  &0.082  &-0.022 &0.145  &-0.122 &-0.149 &0.022  &0.145  &0.092  &0.255\\
32& MoSi$_2$P$_4$ &-0.294 &\textbf{-0.303 }&0.137  &-0.030 &-0.050 &0.141  &-0.210 &-0.202 &0.003  &0.183  &-0.003 &0.141\\
33& WSi$_2$P$_4$  &-0.241 &\textbf{-0.252 }&0.185  &0.021  &0.022  &0.210  &-0.159 &-0.146 &0.070  &0.259  &0.051  &0.192\\
34& CrGe$_2$P$_4$ &0.001  &\textbf{-0.015 }&0.219  &0.197  &0.118  &0.187  &-0.006 &-0.006 &0.152  &0.194  &0.230  &0.305\\
35& MoGe$_2$P$_4$ &-0.127 &\textbf{-0.144 }&0.348  &0.105  &0.087  &0.163  &-0.052 &-0.036 &0.197  &0.197  &0.126  &0.203\\
36& WGe$_2$P$_4$  &-0.077 &\textbf{-0.094 }&0.423  &0.152  &0.168  &0.238  &-0.005 &0.013  &0.259  &0.276  &0.175  &0.249\\
\hline\hline
\end{tabular}
\end{center}
\end{table*}

\newpage
\textcolor{white}{123}
\begin{table*}
\setlength{\tabcolsep}{1.5mm}
\begin{center}
\parbox[c]{18.0cm}{{\bf Table S2,} \normalsize Structural and electronic properties of 36 $MA_{2}Z_{4}$
monalayers with three energetically favorable phases ($\alpha_1$-, $\alpha_2$- and $\beta_2$-phase) extracted
from \textbf{\textcolor{blue}{Table S1}}. Lattice constant
($a$), bond length ($d_{A-Z}^{v}$, $d_{Z-M}$ and $d_{A-Z}^{h}$, where $v$ or $h$ represents the bond along vertical
or horizontal direction), band gap calculated with PBE and HSE06 XC functionals ($E_{g}^{PBE}$ and $E_{g}^{HSE06}$),
respectively, magnets $\mu_{B}$, the type
of energetic favorable phase and dynamical stability, where VEC is the the number of valence
electrons of primitive cell of $MA_2Z_4$ monolayers} \label{Efavorable-36}
\begin{tabular}{cclccccccccccccc}
\hline\hline
VEC & No.&  Name              &   $a$   & $d_{A-Z}^{v}$ & $d_{Z-M}$ & $d_{A-Z}^{h}$ & $E_{g}^{PBE}$  &  $E_{g}^{HSE06}$     & mag   & Phase   & Dynamics       \\
 &  &                    & ({\AA}) & ({\AA}) & ({\AA}) & ({\AA}) & (eV) & (eV)  &  ($\mu_{B}$) &  & (Y/N) \\ \hline
\multirow{12}{*}{32 VEC}
&01 & TiSi$_{2}$N$_{4}$  & 2.95  & 1.77 & 2.04 & 1.75 & 0.61($\Gamma$-M) & 1.60($\Gamma$-M)  &  ---  & $\beta_2$ &  N   \\
&02 & ZrSi$_{2}$N$_{4}$  & 3.05  & 1.83 & 2.16 & 1.75 & 1.00($\Gamma$-M) & 1.98($\Gamma$-M)  &  ---  & $\beta_2$ &  Y   \\
&03 & HfSi$_{2}$N$_{4}$  & 3.04  & 1.82 & 2.14 & 1.75 & 1.21($\Gamma$-M) & 2.21($\Gamma$-M)  &  ---  & $\beta_2$ &  Y   \\
&04 & TiGe$_{2}$N$_{4}$  & 3.08  & 1.89 & 2.08 & 1.89 & 0.82($\Gamma$-M) & 1.86($\Gamma$-M)  &  ---  & $\beta_2$ &  N   \\
&05 & ZrGe$_{2}$N$_{4}$  & 3.19  & 1.93 & 2.19 & 1.89 & 1.04($\Gamma$-$\Gamma$) & 2.34($\Gamma$-$\Gamma$)& --- & $\beta_2$ &  Y   \\
&06 & HfGe$_{2}$N$_{4}$  & 3.18  & 1.93 & 2.17 & 1.89 & 1.15($\Gamma$-$\Gamma$) & 2.45($\Gamma$-$\Gamma$)& --- & $\beta_2$ &  Y   \\
&07 & TiSi$_{2}$P$_{4}$  & 3.53  & 2.27 & 2.49 & 2.22 & ---  & ---  &   ---  & $\beta_2$ &  Y   \\
&08 & ZrSi$_{2}$P$_{4}$  & 3.61  & 2.30 & 2.61 & 2.22 & ---  & ---  &   ---  & $\beta_2$ &  Y   \\
&09 & HfSi$_{2}$P$_{4}$  & 3.61  & 2.30 & 2.59 & 2.22 & ---  & ---  &   ---  & $\beta_2$ &  Y   \\
&10 & TiGe$_{2}$P$_{4}$  & 3.64  & 2.36 & 2.51 & 2.32 & ---  & ---  &   ---  & $\beta_2$ &  Y   \\
&11 & ZrGe$_{2}$P$_{4}$  & 3.72  & 2.38 & 2.63 & 2.32 & ---  & ---  &   ---  & $\beta_2$ &  Y   \\
&12 & HfGe$_{2}$P$_{4}$  & 3.72  & 2.38 & 2.60 & 2.33 & ---  & ---  &   ---  & $\beta_2$ &  Y   \\ \hline
\multirow{12}{*}{33 VEC}
&13 & VSi$_{2}$N$_{4}$   & 2.88  & 1.75 & 2.03 & 1.75 & ---  & ---  &   0.97 & $\alpha_1$ &  Y   \\
&14 & NbSi$_{2}$N$_{4}$  & 2.96  & 1.78 & 2.13 & 1.75 & ---  & ---  &   0.57 & $\alpha_1$ &  Y   \\
&15 & TaSi$_{2}$N$_{4}$  & 2.97  & 1.78 & 2.13 & 1.75 & ---  & ---  &   ---  & $\alpha_1$ &  Y   \\
&16 & VGe$_{2}$N$_{4}$   & 3.05  & 1.87 & 2.06 & 1.89 & ---  & ---  &   0.98 & $\beta_2$  &  N   \\
&17 & NbGe$_{2}$N$_{4}$  & 3.09  & 1.89 & 2.16 & 1.90 & ---  & ---  &   0.72 & $\alpha_1$ &  Y   \\
&18 & TaGe$_{2}$N$_{4}$  & 3.08  & 1.87 & 2.15 & 1.88 & ---  & ---  &   0.49 & $\alpha_1$ &  Y   \\
&19 & VSi$_{2}$P$_{4}$   & 3.48  & 2.25 & 2.43 & 2.25 & 0.00 & ---  &   1.00 & $\alpha_1$ &  Y   \\
&20 & NbSi$_{2}$P$_{4}$  & 3.53  & 2.27 & 2.52 & 2.23 & ---  & ---  &   ---  & $\alpha_1$ &  Y   \\
&21 & TaSi$_{2}$P$_{4}$  & 3.54  & 2.27 & 2.52 & 2.24 & ---  & ---  &   ---  & $\alpha_1$ &  Y   \\
&22 & VGe$_{2}$P$_{4}$   & 3.56  & 2.33 & 2.44 & 2.36 & ---  & ---  &   1.00 & $\alpha_2$  &  Y  \\
&23 & NbGe$_{2}$P$_{4}$  & 3.62  & 2.35 & 2.53 & 2.36 & ---  & ---  &   ---  & $\alpha_2$  &  Y  \\
&24 & TaGe$_{2}$P$_{4}$  & 3.61  & 2.34 & 2.53 & 2.36 & ---  & ---  &   ---  & $\alpha_2$  &  Y  \\ \hline
\multirow{12}{*}{34 VEC}
&25 & CrSi$_{2}$N$_{4}$  & 2.84  & 1.73 & 2.00 & 1.75 & 0.49($\Gamma$-K)  & 0.94(K-K)         &  ---  & $\alpha_1$ &  Y   \\
&26 & MoSi$_{2}$N$_{4}$  & 2.91  & 1.75 & 2.09 & 1.75 & 1.74($\Gamma$-K)  & 2.31($\Gamma$-K)  &  ---  & $\alpha_1$ &  Y   \\
&27 & WSi$_{2}$N$_{4}$   & 2.91  & 1.76 & 2.10 & 1.75 & 2.08($\Gamma$-K)  & 2.66($\Gamma$-K)  &  ---  & $\alpha_1$ &  Y   \\
&28 & CrGe$_{2}$N$_{4}$  & 3.06  & 1.88 & 2.04 & 1.89 & ---  & ---  &  2.00    & $\beta_2$ &  N   \\
&29 & MoGe$_{2}$N$_{4}$  & 3.02  & 1.85 & 2.12 & 1.87 & 0.99($\Gamma$K-K) & 1.38($\Gamma$K-K) &  ---  & $\alpha_1$ &  Y   \\
&30& WGe$_{2}$N$_{4}$    & 3.02  & 1.85 & 2.13 & 1.88 & 1.29($\Gamma$K-K)  & 1.69($\Gamma$K-K)&  ---  & $\alpha_1$ &  Y   \\
&31 & CrSi$_{2}$P$_{4}$  & 3.41 & 2.23 & 2.37 & 2.27 & 0.34($\Gamma$K-K)  & 0.65(K-K)         &  ---  & $\alpha_2$  &  Y  \\
&32 & MoSi$_{2}$P$_{4}$  & 3.46 & 2.25 & 2.46 & 2.26 & 0.91(K-K)          & 1.19(K-K)         &  ---  & $\alpha_2$  &  Y  \\
&33 & WSi$_{2}$P$_{4}$   & 3.46 & 2.25 & 2.46 & 2.26 & 0.86(K-K)          & 1.11(K-K)         &  ---  & $\alpha_2$  &  Y  \\
&34 & CrGe$_{2}$P$_{4}$  & 3.49 & 2.31 & 2.39 & 2.36 & 0.04($\Gamma$K-K)  & 0.36($\Gamma$K-K) &  ---  & $\alpha_2$  &  Y  \\
&35 & MoGe$_{2}$P$_{4}$  & 3.53  & 2.32 & 2.47 & 2.34 & 0.56($\Gamma$K-K) & 0.95($\Gamma$K-K) &  ---  & $\alpha_2$  &  Y  \\
&36 & WGe$_{2}$P$_{4}$   & 3.54  & 2.32 & 2.47 & 2.35 & 0.63($\Gamma$K-K) & 0.89(K-K)         &  ---  & $\alpha_2$  &  Y  \\
\hline\hline
\end{tabular}
\end{center}
\end{table*}

\newpage
\textcolor{white}{123}
\begin{table*}
\setlength{\tabcolsep}{1.2mm}
\begin{center}
\parbox[c]{18.0cm}{{\bf Table S3,} \normalsize The enthalpies of formation
E$_f$ (eV/atom) of $\alpha_{1-6}$- and $\beta_{1-6}$-$MA_2Z_4$ monolayers
with space group P\={6}m2 and P\={3}m1, respectively, where $M$ = Mg, Ca,
Sr, Zn, Cd, Hg; $A$ = Al and Ga; $Z$ = S, Se and Te. It should be noticed that
because the difference in E$_f$ between $\beta_1$- and $\beta_2$-phase
is very small, we chose four significant digits to distinguish them.} \label{formation-Mg}
\begin{tabular}{clcccccccccccc}
\hline\hline
No.&  Name & $\alpha_1$ & $\alpha_2$ & $\alpha_3$ & $\alpha_4$ & $\alpha_5$ & $\alpha_6$ & $\beta_1$& $\beta_2$& $\beta_3$& $\beta_4$& $\beta_5$& $\beta_6$ \\\hline
01& MgAl$_2$S$_4$  &-1.102 &-1.101 &-1.101 &-0.856 &-1.094 &-0.853 &-1.1525 &\textbf{-1.1542 }&-0.782 &-1.037 &-1.042 &-0.784 \\
02& CaAl$_2$S$_4$  &-1.278 &-1.275 &-1.250 &-0.998 &-1.240 &-0.989 &-1.2891 &\textbf{-1.2931 }&-0.512 &-1.219 &-1.227 &-0.530 \\
03& SrAl$_2$S$_4$  &-1.229 &-1.226 &-1.183 &-0.923 &-1.176 &-0.914 &-1.2290 &\textbf{-1.2329 }&-0.448 &-1.170 &-1.176 &-0.919 \\
04& ZnAl$_2$S$_4$  &-0.820 &-0.821 &-0.827 &-0.591 &-0.831 &-0.604 &\textbf{-0.8854 }&-0.8843 &-0.519 &-0.768 &-0.763 &-0.506 \\
05& CdAl$_2$S$_4$  &-0.810 &-0.812 &-0.792 &-0.559 &-0.805 &-0.573 &\textbf{-0.8398 }&-0.8375 &-0.523 &-0.774 &-0.763 &-0.511 \\
06& HgAl$_2$S$_4$  &-0.623 &-0.628 &-0.614 &-0.417 &-0.635 &-0.431 &\textbf{-0.6474 }&-0.6429 &-0.397 &-0.615 &-0.593 &-0.387 \\
07& MgGa$_2$S$_4$  &-0.804 &-0.803 &-0.719 &-0.513 &-0.715 &-0.514 &-0.8487 &\textbf{-0.8504 }&-0.439 &-0.656 &-0.657 &-0.438 \\
08& CaGa$_2$S$_4$  &-0.974 &-0.972 &-0.881 &-0.431 &-0.867 &-0.653 &-0.9827 &\textbf{-0.9852 }&-0.494 &-0.845 &-0.855 &-0.491 \\
09& SrGa$_2$S$_4$  &-0.925 &-0.923 &-0.823 &-0.447 &-0.812 &-0.588 &-0.9248 &\textbf{-0.9273 }&-0.475 &-0.804 &-0.814 &-0.439 \\
10& ZnGa$_2$S$_4$  &-0.526 &-0.526 &-0.449 &-0.256 &-0.459 &-0.278 &-0.5863 &\textbf{-0.5869 }&-0.195 &-0.393 &-0.379 &-0.174 \\
11& CdGa$_2$S$_4$  &-0.513 &-0.514 &-0.421 &-0.224 &-0.439 &-0.247 &\textbf{-0.5400 }&-0.5398 &-0.194 &-0.405 &-0.387 &0.032 \\
12& HgGa$_2$S$_4$  &-0.330 &-0.331 &-0.249 &0.042  &-0.276 &-0.106 &\textbf{-0.3521 }&-0.3510 &-0.064 &-0.251 &-0.221 &0.207 \\
13& MgAl$_2$Se$_4$ &-0.907 &-0.906 &-0.899 &-0.674 &-0.888 &-0.673 &-0.9531 &\textbf{-0.9546 }&-0.602 &-0.835 &-0.841 &-0.602 \\
14& CaAl$_2$Se$_4$ &-1.103 &-1.100 &-1.083 &-0.852 &-1.065 &-0.838 &-1.1113 &\textbf{-1.1152 }&-0.493 &-1.042 &-1.057 &-0.520 \\
15& SrAl$_2$Se$_4$ &-1.073 &-1.069 &-1.044 &-0.807 &-1.028 &-0.793 &-1.0720 &\textbf{-1.0762 }&-0.463 &-1.018 &-1.032 &-0.458 \\
16& ZnAl$_2$Se$_4$ &-0.653 &-0.654 &-0.655 &-0.445 &-0.655 &-0.460 &\textbf{-0.7164 }&-0.7157 &-0.386 &-0.595 &-0.588 &-0.373 \\
17& CdAl$_2$Se$_4$ &-0.664 &-0.665 &-0.649 &-0.441 &-0.657 &-0.457 &\textbf{-0.6942 }&-0.6929 &-0.407 &-0.624 &-0.613 &-0.393 \\
18& HgAl$_2$Se$_4$ &-0.507 &-0.510 &-0.506 &0.127  &-0.523 &-0.353 &\textbf{-0.5332 }&-0.5302 &-0.325 &-0.499 &-0.476 &-0.324 \\
19& MgGa$_2$Se$_4$ &-0.702 &-0.701 &-0.610 &-0.418 &-0.602 &-0.423 &-0.7428 &\textbf{-0.7445 }&-0.358 &-0.549 &-0.550 &-0.354 \\
20& CaGa$_2$Se$_4$ &-0.889 &-0.887 &-0.799 &-0.444 &-0.779 &-0.583 &-0.8943 &\textbf{-0.8971 }&-0.407 &-0.756 &-0.771 &-0.507 \\
21& SrGa$_2$Se$_4$ &-0.857 &-0.854 &-0.767 &-0.460 &-0.747 &-0.543 &-0.8554 &\textbf{-0.8584 }&-0.462 &-0.737 &-0.464 &-0.452 \\
22& ZnGa$_2$Se$_4$ &-0.452 &-0.452 &-0.373 &-0.206 &-0.380 &-0.231 &-0.5131 &\textbf{-0.5138 }&-0.167 &-0.321 &-0.302 &-0.156 \\
23& CdGa$_2$Se$_4$ &-0.460 &-0.459 &-0.369 &-0.075 &-0.383 &-0.222 &-0.4882 &\textbf{-0.4886 }&-0.173 &-0.348 &-0.329 &-0.160 \\
24& HgGa$_2$Se$_4$ &-0.307 &-0.307 &-0.237 &0.011  &-0.260 &-0.118 &-0.3330 &\textbf{-0.3331 }&-0.077 &-0.233 &-0.203 &0.104 \\
25& MgAl$_2$Te$_4$ &-0.480 &-0.480 &-0.454 &-0.246 &-0.442 &-0.254 &-0.5252 &\textbf{-0.5264 }&-0.203 &-0.394 &-0.394 &-0.202 \\
26& CaAl$_2$Te$_4$ &-0.695 &-0.692 &-0.667 &-0.443 &-0.640 &-0.428 &-0.7012 &\textbf{-0.7047 }&-0.239 &-0.618 &-0.638 &-0.284 \\
27& SrAl$_2$Te$_4$ &-0.688 &-0.685 &-0.661 &-0.156 &-0.634 &-0.415 &-0.6866 &\textbf{-0.6907 }&-0.231 &-0.623 &-0.644 &-0.255 \\
28& ZnAl$_2$Te$_4$ &-0.263 &-0.264 &-0.263 &-0.097 &-0.269 &-0.113 &\textbf{-0.3315 }&-0.3307 &-0.066 &-0.214 &-0.199 &-0.071 \\
29& CdAl$_2$Te$_4$ &-0.300 &-0.301 &-0.279 &-0.108 &-0.289 &-0.120 &\textbf{-0.3366 }&-0.3357 &-0.088 &-0.254 &-0.235 &-0.094 \\
30& HgAl$_2$Te$_4$ &-0.185 &-0.187 &-0.202 &0.269  &-0.214 &-0.069 &\textbf{-0.2184 }&-0.2162 &-0.057 &-0.189 &-0.184 &-0.065 \\
31& MgGa$_2$Te$_4$ &-0.372 &-0.371 &-0.267 &-0.075 &-0.267 &-0.130 &-0.4134 &\textbf{-0.4148 }&-0.114 &-0.226 &-0.219 &-0.107 \\
32& CaGa$_2$Te$_4$ &-0.577 &-0.575 &-0.473 &-0.299 &-0.450 &-0.281 &-0.5801 &\textbf{-0.5829 }&-0.292 &-0.276 &-0.444 &-0.265 \\
33& SrGa$_2$Te$_4$ &-0.567 &-0.564 &-0.471 &-0.319 &-0.444 &-0.314 &-0.5641 &\textbf{-0.5674 }&-0.296 &-0.281 &-0.453 &-0.308 \\
34& ZnGa$_2$Te$_4$ &-0.161 &-0.161 &-0.099 &0.116  &-0.111 &-0.001 &-0.2267 &\textbf{-0.2269 }&0.016  &-0.059 &-0.056 &0.031 \\
35& CdGa$_2$Te$_4$ &-0.193 &-0.192 &-0.110 &0.058  &-0.126 &-0.003 &-0.2279 &\textbf{-0.2284 }&0.022  &-0.092 &-0.081 &0.026 \\
36& HgGa$_2$Te$_4$ &-0.083 &-0.082 &0.170  &0.131  &-0.056 &0.056  &\textbf{-0.1144} &-0.1143 &0.080  &-0.025 &-0.024 &0.073 \\ \hline\hline
\end{tabular}
\end{center}
\end{table*}

\newpage
\textcolor{white}{123}
\begin{table*}
\setlength{\tabcolsep}{4.0mm}
\begin{center}
\parbox[c]{17.0cm}{{\bf Table S4,} \normalsize Carrier mobility of $\alpha_2$-WSi$_{2}$P$_{4}$
,2$H$-MoS$_2$ and 2$H$-WSe$_2$. Deformation potential $E_{1}$ (eV), in-plane stiffness $C^{2D}$ (N/m),
effective mass $m^{*}$ ($m_{e}$), mobility $\mu$ (cm$^{2}$ V$^{-1}$ s$^{-1}$) for electron (e) and hole (h) along a$_{o1}$ (or zigzag)
and a$_{o2}$ (or armchair) directions (See \textbf{\textcolor{blue}{Figure S4}}) at 300 K.} \label{table-cmm}
\begin{tabular}{cccccc}
\hline\hline
  Compounds                 &  carrier type   & $E_{1}$ & $C^{2D}$  & $m^{*}$   &  $\mu$  \\
\multirow{4}{*}{$\alpha_2$-WSi$_{2}$P$_{4}$ (this work)} &   e(a$_{o1}$)(K) & -10.90      & 227.49     & 0.43 &  147.22 \\
                                              &   h(a$_{o1}$)(K)          & -6.75       & 227.49     & 0.35 &  466.70 \\
                                              &   e(a$_{o2}$)(K)          & -10.71      & 227.74     & 0.43 &  152.66 \\
                                              &   h(a$_{o2}$)(K)          & -6.82       & 227.74     & 0.35 &  457.67 \\ \hline
 \multirow{4}{*}{MoS$_{2}$ (this work)}       &   e(a$_{o1}$)(K)          & -11.24      & 132.04     & 0.46 &   70.22 \\
                                              &   h(a$_{o1}$)(K)          & -5.68       & 132.04     & 0.59 &  169.45 \\
                                              &   e(a$_{o2}$)(K)          & -11.25      & 132.74     & 0.46 &   70.37 \\
                                              &   h(a$_{o2}$)(K)          & -5.70       & 132.74     & 0.59 &  168.91 \\ \hline
 \multirow{4}{*}{WSe$_{2}$ (this work)}       &   e(a$_{o1}$)(K)          & -10.95      & 119.52     & 0.34 &  119.76 \\
                                              &   h(a$_{o1}$)(K)          & -4.92       & 119.52     & 0.47 &  321.87 \\
                                              &   e(a$_{o2}$)(K)          & -10.29      & 119.77     & 0.35 &  130.53 \\
                                              &   h(a$_{o2}$)(K)          & -5.03       & 119.77     & 0.47 &  309.92 \\ \hline
 \multirow{4}{*}{MoS$_{2}$ (Ref ~\citenum{CaiZhang2014})} &   e(a$_{o1}$)(K)  & -10.88      & 127.44     & 0.46 &   72.16 \\
                                              &   h(a$_{o1}$)(K)          & -5.29       & 127.44     & 0.57 &  200.52 \\
                                              &   e(a$_{o2}$)(K)          & -11.36      & 128.16     & 0.48 &   60.32 \\
                                              &   h(a$_{o2}$)(K)          & -5.77       & 128.16     & 0.60 &  152.18 \\ \hline
 \multirow{4}{*}{WSe$_{2}$ (Ref ~\citenum{zhuang_CM_2013})} &   e(a$_{o1}$)(K)          & -10.23     & 121.10     & 0.35 &  135.08 \\
                                              &   h(a$_{o1}$)(K)          & -4.65       & 121.10     & 0.46 &  373.65 \\
                                              &   e(a$_{o2}$)(K)          & -10.71      & 120.80    & 0.33 &   --- \\
                                              &   h(a$_{o2}$)(K)          & -4.52       & 120.80     & 0.44 &  --- \\

\hline\hline
\end{tabular}
\end{center}
\end{table*}

\end{document}